\documentclass[epj]{svjour}
\onecolumn
\usepackage{graphics}
\begin{document}
\title{A Theoretical Study of the Magnetically Deformed Inner Crust Matter 
of Magnetars}
\author{Arpita Ghosh and Somenath Chakrabarty$^\dagger$}
\institute{Department of Physics, Visva-Bharati, Santiniketan, India
741 235 \\$^\dagger$E-mail:somenath.chakrabarty@visva-bharati.ac.in }
\date{Received:{\today} / Revised version: date}
\abstract{
We have studied various physical properties of magnetically deformed atoms and the associated matter, 
replacing the atoms by the deformed Wigner-Seitz (WS) cells at the crustal region of strongly magnetized
neutron stars (magnetars). A relativistic version of Thomas-Fermi (TF) model in presence of strong magnetic
field in cylindrical coordinates is used to study the properties of such matter.
}
\maketitle
\section{Introduction}
From the observational evidence of a few strongly magnetized neutron stars, which are supposed to be the 
sources of
anomalous X-rays and soft gamma rays, also called magnetars \cite{R1,R2,R3,R4}, the study of the
effect of strong magnetic field on dense neutron star matter, including the crustal matter, 
both outer crust and inner crust regions of such compact stellar objects have gotten a new dimension. 
These exotic 
objects are also called anomalous X-ray pulsars (AXP) and soft gamma repeaters (SGR). The outer crust of a 
typical neutron star in general, is mainly composed of dense crystalline metallic iron \cite{hpy}. 
The density of
the upper edge of such metallic crystalline matter 
is $\sim 10^6-10^7$gm cm$^{-3}$, whereas the bottom edge is $\sim 10^{11}$gm cm$^{-3}$, the matter
consists of nuclei (also some highly neutron rich nuclei), surrounded by cylindrically deformed
distribution of electron gas, makes
the system electrically charge neutral and 
drifted out free neutrons, if the density of matter is $\geq$ neutron drift value. Since the density is 
high eneough,  
it is therefore absolutely impossible 
to investigate the properties of such matter in material
science laboratories, even at zero magnetic field. The observed surface magnetic field of the magnetars
is $\sim 10^{15}$G, which is again too high to achieve in the terrestrial laboratories. Also, it is quite 
possible that the interior field of such exotic objects can go up to $\sim 10^{18}$G (which can be
shown theoretically by virial theorem). If the magnetic field at the interior is really so high, then most 
of the physical and chemical properties of dense neutron matter should change significantly from the 
conventional neutron star (radio
pulsar) scenario (see the recent article by one of the co-authors \cite{R5} for necessary 
references). In \cite{R5}, the matter in the outer crust region of strongly
magnetized neutron stars have been studied using Thomas-Fermi approximation, in which the WS cells are assumed
to be spherical in nature and arranged in a regular manner to form dense crystal of metallic iron ($\sim$
BCC-type). 
Even though the magnetic field strength at the crustal region is slightly higher 
than  $10^{15}$G, it must change significantly most of the properties of dense matter, 
both in the outer crust and 
the inner crust regions of the magnetars \cite{R5,R6}. It is believed that
strong magnetic field can cause a structural deformation of the metallic atoms (see e.g., \cite{OST}) 
present in the inner
crust region of a neutron star. The spherically symmetric structure of the atoms are destroyed and become 
cigar shape with 
the elongated axis along the direction of strong magnetic field. The atoms may even become almost an one 
dimensional string like object, i.e., needle shape, if the magnetic field strength is extremely high. In
presence of ultra-strong magnetic field of strength $>4.4\times 10^{13}$G, the application of TF model for
spherically symmetric WS cells is not a valid approximation \cite{LIMOD}. However, one can use TF model
for sylindrically deformed WS cells because of ultra-strong magnetic field \cite{OST,BKS,XUE1,XUE2}. In
presence of ultra-strong magnetic field ($\gg 4.4\times 10^{13}$G), the WS cells get magnetically
deformed and become ellipsoidal in nature. 
In this article, for the sake of simplicity, we shall assume a cylindrical type deformation of the atoms 
in the inner crust region and use cylindrical coordinate system with azimuthal symmetry. 
In reality, to investigate the cigar like deformed atoms in presence of strong magnetic field 
one has to use prolate spheroidal coordinate system \cite{RR6}.  
In future we shall present the problem related to structural deformation of atoms 
in a strong quantizing magnetic field using such coordinate system \cite{RR6}.
                                           
In the present article we shall study the properties of inner
crust matter composed of magnetically deformed metallic atoms. In section 2 we have developed
the basic formalism and discuss the numerical results, whereas in the last section we
have given the conclusions and the future prospect of this work.
\section{Basic formalism}
The width of the outer crust of a typical neutron star is $\sim 0.3$ km, the density of matter, which is
assumed to be a dense crystalline structure  of metallic iron is $\sim 10^6-10^{11}$gm cm$^{-3}$, ranging
from the upper
edge to the bottom edge \cite{SHA} respectively. In one of our previous work appeared i1n \cite{R5}, 
the properties of
outer crust region has been studied. To investigate the
properties of such dense exotic crystalline matter of metallic iron, we have
replaced the outer crust matter by a regular array of spherically symmetric  WS cells, 
with positively charged nuclei at the centre surrounded by a non-uniform electron gas. Whereas in the
inner crust region, of width $\sim 0.5-0.7$km and density $\sim 10^{11}-10^{14}$gm cm$^{-3}$, we assume 
that since the magnetic field is high enough compared to outer crust region, the electron distribution
around each nucleus (iron and also some neutron rich nuclei) gets deformed (the numerical values of widths
and density ranges for various regions inside a neutron star strongly depends on the type of equation of
state considered). They become cigar shape. In
\cite{R6} we have studied the equation of states of inner crust matter with spherically symmetric electron
distribution around each nucleus inside WS cells in presence of strong magnetic field. In this article
for the sake of
simplicity, we assume cylindrical type distribution of electron gas around each nucleus with the axis of
each cylinder is along the direction of magnetic field and further assume azimuthal symmetry for all the
cylindrically deformed WS cells. In this article, although we have considered magnetically deformed WS
cells for electron distribution, the nuclei at the centre of these cells are assumed to be spherical
in nature. We have also assumed that the magnetic field is not so strong to populate proton Landau levels and
the magnetic (dipole) energy of neutron sector is negligibly small compared to the kinetic energy of these
particles. 

To investigate various physical properties of deformed electron distribution in the inner crust matter, we start 
with the Poisson's equation, given by
\begin{equation}
\nabla^2\phi=4\pi en_e \theta(\vec{r}-\vec{r_n})-\frac{4\pi Ze}{V_n}\theta(\vec{r_n}-\vec{r})
\end{equation}
where $V_n=4\pi r_n^3/3=4\pi r_0^3A/3$, is the nuclear volume, $r_0=1.12$fm and $A$ is the mass number of
the nucleus \cite{XUE1,XUE2}.
In the cylindrical coordonate system, if one assumes the magnetic deformation of atomic nuclei in this
region, the $\theta$-function associated with the  contribution of protons within the nucleus has to be
replaced by $\theta(\vec{r_n}- \vec{r})\theta(z_n-z)$, where in eqn.(1), $Z$ is the atomic number of the 
nucleus,
$r_n$ is the nuclear radius, assumed to be spherical in shape, whereas in the above text, $r_n$ and $z_n$ 
given in the
arguments of the $\theta$-functions are respectively the radial and axial dimensions of the cylidrically
deformed nucleus. However, in this article we have not assumed magnetic deformation of the nuclei.
Here $\phi$ is the electrostatic field, $e$ is the electron charge and $n_e$ is the electron density,
which because of assumed non-uniformity within the WS cells,  is a function of positional coordinates $(r,z)$.
In our study, we shall consider the variation of $\phi(r,z)$ within the cylindrical distribution of the
electrons surrounding the positively charged nucleus. Therefore, in the Poisson's
equation the nuclear part as shown in eqn.(1) will not contribute. 
Now in the cylindrical coordinate with circular symmetry, the above equation reduces to
\begin{equation}
\frac{\partial^2\phi}{\partial r^2}+\frac{1}{r}\frac{\partial\phi}{\partial
r}+\frac{\partial^2\phi}{\partial z^2}=4\pi en_e
\end{equation}
It is well known that in presence of strong quantizing magnetic field, the number density of
degenerate electron gas is given by 
\begin{equation}
n_e=\frac{eB}{2\pi^2}\sum_{\nu=0}^{\nu_{max}}(2-\delta_{\nu 0})p_F
\end{equation}
where B is the constant external magnetic field, assumed to be acting along z-direction and is $>B_c
^{(e)}$, where $B_c
^{(e)}$ is the typical strength of magnetic field at and above which, in the relativistic scenario the Landau
levels for the electrons are populated. For the sake of convenience, throughout this article we shall
use natural units, i.e., $\hbar=c=1$. The critical strength is then given by $B_c=m_e ^2/\mid e\mid$ \cite{R5},
where $m_e$ is the electron rest mass and $\mid e\mid$ is the magnitude of electron charge. This critical
strength may be obtained by equating the cyclotron quantum with the rest mass energy for electrons.
We further assume that the matter is at zero temperature.
In eqn.(3), $p_F$ is the electron Fermi momentum, $\nu$ is the Landau quantum
number, with $\nu_{max}$, the upper limit of $\nu$. The upper limit will be finite at zero temperature
and infinity for finite temperature. The factor $(2-\delta_{\nu 0})$ takes care of
singly degenerate $\nu=0$ state and doubly degenerate all other states with $\nu\neq 0$.
To study the properties of inner crust matter with deformed WS cells, we make Thomas-Fermi
approximation, which is  semi-classical version of Hartree approximation
\cite{R7}. In this model, the well known Thomas-Fermi condition is given by
\begin{equation}
\mu_e= (p_F ^2+m_e ^2+2\nu eB)^{1/2}-e\phi={\rm{constant}}
\end{equation}
where $\mu_e$ is the electron chemical potential, which is assumed to be constant
throughout the WS cell.
Hence one can express the Fermi momentum for electrons in the following form: 
\begin{equation}
p_F=[(\mu_e+e\phi)^2-m_e ^2-2\nu eB]^{1/2}
\end{equation}
Since the electrostatic potential $\phi\equiv \phi(r,z)$, the Fermi momentum $p_F$ for the electron 
is also a function of positional coordinates $(r,z)$ within the cell. In principle
one should use this exact expression for electron Fermi momentum in the equation for 
electron density (eqn.(3)) which in turn appearing on the right hand side of the
cylindrical form of Poisson's equation (eqn.(2)). However, with this exact expression for $p_F$, since
the equation becomes non-linear in $\phi(r,z)$, it is
absolutely impossible to proceed further analytically, even a single effective step. From the very
beginning, therefore, one has to use some numerical technique to solve the Poisson's equation. Of course, 
with the numerical method 
within the limitation of the algorithm followed, more exact results can be obtained.
However, in numerical computation of $\phi(r,z)$, we only get a set of numbers, but the beauty of this model
will be completely destroyed and a lot of interesting physics associated with the intermediate results of 
this problem will be totally lost. Therefore, to get an approximate analytical
solution for $\phi(r,z)$, as a first approximation, we set the upper limit for Landau quantum
number $\nu_{max}=0$ and also neglect the rest mass of electrons, i.e., we put $m_e=0$ (since in the inner
crust region the density
of electron gas  is high enough, the electron Fermi momentum will also be large compared to electron rest
mass, therefore we expect that without appreciable error one can neglect electron mass in the above expression)
in the expression for Fermi momentum $p_F$ (eqn.(5)).  The approximation $\nu_{max}=0$ is actually valid if
the magnetic field is extremely high ($\sim 10^{15}$G), perhaps is a valid approximation at the inner
crust region for magnetars. 
However, to investigate some of the properties of dense electron gas
within the cylindrically deformed WS cells to somewhat exact manner, later in this article, we
shall use this approximate solution for $\phi(r,z)$ only, but do not use the first
approximation $\nu_{max}=0$ and $m_e=0$ to evaluate  mathematical expressions for various physical
quantities of the dense electron gas. 
However. later in this article we shall show that the upper limit 
$\nu_{max}$ for the electron Landau quantum number is also a function of $(r,z)$. 
Therefore to evaluate various
physical quantities analytically in this region, first, one has to obtain  an approximate 
solution for the Poisson's equation (with the values of $\nu_{max}=0$ and $m_e=0$ in the first
approximation). To achieve our objectives, we use first the approximate form of electron Fermi momentum
obtained from the assumption as mentioned above and its mathematical form is  given by
\begin{equation}
p_F \approx \mu_e+e\phi
\end{equation}
Next on substituting
\begin{equation}
\mu_e+e\phi(r,z)=\psi(r,z),
\end{equation}
the cylindrical form of Poisson's equation reduces to
\begin{equation}
\frac{\partial^2\psi}{\partial r^2}+\frac{1}{r}\frac{\partial\psi}{\partial
r}+\frac{\partial^2\psi}{\partial z^2}=\lambda^2\psi
\end {equation}
where
$\lambda^2=2e^3 B/\pi$. From eqn.(8), it is quite obvious that under this approximation, 
the Poisson's equation reduces to a linear partial differential equation.  
To solve this  equation analytically we 
use the method of separation of variables, given by 
\begin{equation}
\psi(r,z)=R(r)L(z)
\end{equation} 
Substituting $\psi(r,z)$ from eqn.(9) in eqn.(8) and introducing a constant $\xi$, we get
\begin{eqnarray}
&&\frac{d^2 R}{dr^2}+\frac{1}{r}\frac{dR}{dr}+\xi^2 R=0\\
&&\frac{d^2 L}{dz^2}-(\xi^2+\lambda^2)L=0
\end{eqnarray}
where $\xi$ is some real constant, independent of r and z but may change with the magnetic field
strength and with the mass number and the atomic number of the type of elements present in this
region. The solutions of eqns.(10) and (11) are well known. For eqn.(10), the solution is an ordinary  Bessel
function of order zero with the argument $\xi r$, whereas for eqn.(11), it is an exponentially
decaying function of $z$. In the language of mathematics,    
the solution for $\psi(r,z)$ is then given by
\begin{equation}
\psi(r,z)=CJ_0(\xi r)exp\left [\pm(\xi^2+\lambda^2)^{1/2}z\right ]
\end{equation}
where $+$ and $-$ signs are for $z<0$ or $>0$ respectively. We consider a convenient form of
cylindrical coordinate system, such that $z=0$ plane is at the middle of the finite size cylinder. 
In that case we have to take positive sign for the upper half of the cylinder and negative sign 
for the lower half. 
Here $C$ is a constant (again may change with 
the magnetic field strength and with
the nuclear properties of the elements present in the inner crust region) and $J_0(\xi r)$
is the Bessel function of order zero.
Now on the nuclear surface, at the centre of the WS cells, $\phi=Z e/r_n $ \cite{R9}, where $Z$ is the 
atomic number and $r_n=r_0 A^{1/3}$ is the nuclear radius, $r_0=1.12$fm and $A$ is the mass number. 
We consider eqn.(12) at various point on the nuclear 
surface. To evaluate the parameters $C$ and $\xi$ numerically, we do the following: put $r=\alpha r_n$ and 
$z=\beta r_n$ at a particular point on the nuclear surface and also consider $r=\beta r_n$ and $z=\alpha r_n$
for another point, with $\alpha^2+\beta^2=1$. Then we have
\begin{equation}
\psi(\alpha r_n,\beta r_n)=CJ_0(\alpha r_n\xi)\exp \left [-(\xi^2+\lambda^2)^{1/2}\beta r_n \right ]=\mu_e+\frac{Ze^2}{r_n}
\end{equation}
and
\begin{equation}
\psi(\beta r_n,\alpha r_n)=CJ_0(\beta r_n\xi)\exp \left [-(\xi^2+\lambda^2)^{1/2}\alpha r_n \right
]=\mu_e+\frac{Ze^2}{r_n}
\end{equation}
From eqn.(14) we have
\begin{equation}
C=\frac{1}{J_0(\beta r_n\xi)\exp \left [-(\xi^2+\lambda^2)^{1/2}\alpha r_n\right ]}\left [\mu_e+\frac{Ze^2}{r_n}\right ]
\end{equation}
Combining eqns.(13)-(14) we get 
\begin{equation}
J_0(\alpha r_n\xi)\exp \left [-(\xi^2+\lambda^2)^{1/2}\beta r_n \right ]=J_0(\beta r_n\xi)\exp \left
[-(\xi^2+\lambda^2)^{1/2}\alpha r_n \right ]
\end{equation}
Hence we can write 
\begin{equation}
\exp\left [-(\xi^2 +\lambda^2)^{1/2}r_n (\beta-\alpha)\right ]=\frac{J_0(\beta r_n\xi)}{J_0(\alpha
r_n\xi)}
\end{equation}
This is a  highly transcendental equation for $\xi$. However, it is possible to evaluate $\xi$ numerically 
from this equation for a given magnetic field strength and for a given type of element, e.g., for metallic 
iron.
To obtain $\xi$ numerically we express eqn.(17) in the 
following convenient form
\begin{equation}
\xi^2+\lambda^2=\frac{1}{r_n^2(\beta-\alpha)^2}\left [\ln \left \{\frac{J_0(\beta r_n\xi)}{J_0(\alpha r_n\xi)}\right\} \right
]^2
\end{equation}
To evaluate numerical values for $\xi$, we now put
(a) $r$ and $z$ values of the equator and poles of the nucleus, (b) $r=r_n/4$, (c) $r=r_n/2$ and (d)
$r=3r_n/4$, for four different positions on the nuclear surface and obtain four different sets of
$\xi$ as a function of magnetic field strength. For all these cases the $z$-coordinates are obtained 
from the relation $z=(r_n^2-r^2)^{1/2}$. 
Because of the symmetry about $z=0$ plane, the choice of negative values for $\alpha$ and $\beta$ will
give identical results. 
In fig.(1) we have plotted $\xi$ (in MeV) as a function of magnetic field strength $B$, expressed in terms of
critical magnetic field strength $B_c^{(e)}$, for the specific values of $r$ and $z$ as indicated above 
by (a), (b), (c) and (d). For all these cases, the variations are insensitive for the low
and moderate values of magnetic field strengths. This figure shows that beyond field strength $10^{17}$G, 
$\xi$ increases sharply for all the cases. It is also obvious from fig.(1) that the parameter $\xi$
not only varies with the strength of magnetic field but also more strongly depends on the positional
coordinates on the nuclear surface, giving one of the boundary conditionss. 
In fig.(2) we have plotted the same kind of variations for the parameter $C$ with the magnetic field
strength, expressed in the same unit as in fig.(1) and also considering the same positions on the nuclear
surface as considered for fig.(1).
For low and moderate magnetic field values it is also almost constant, then it falls abruptly beyond
$10^{17}$G and finally saturates to some constant values. Since beyond $10^{17}$G,
electrons within the cells occupy their zeroth or very low lying Landau levels, the quantum mechanical
effect of magnetic field becomes extremely important in this region and as a consequence both $\xi$ 
and $C$ change significantly beyond this magnetic field value. 
Since the magnetic field is not ultra strong to distort atomic
nuclei, we have therefore considered spherical nuclei of radii $r_0A^{1/3}$, at
the $z$ and $r$ symmetric position inside the cylinder.
From the solutions of eqns.(13) and (14), we have noticed that the functional form of the solution given
by eqn.(12)
does not change, but the unknown parameters $C$ and $\xi$ can have a large number of
roots. As a result the final solution of Poisson's equation becomes degenerate (which has been shown in
figs.(1) and (2)). 
As an hypothetical case, instead of spherically
symmetric nuclei, we have replaced them by cylindrically deformed nuclei with axially symmetric nucleon
distribution. The smaller cylinder (deformed nucleus) is coaxial with the bigger one and with identical 
$z=0$ plane.
For the surface potential we have used eqn.(52) of Appendix A. In eqn.(52) we have replaced 
$r_{max}$ and $z_{max}$ by
$r_n$ and $z_n$, the maximum values of $r$ and $z$ for the deformed nuclei. Here $r_1$ and $z_1$ are the
surface values of $r$ and $z$ coordinates at any arbitrary point on the nuclear surface. To make the nucleon
distributions within the nuclei axially symmetric, we put $\theta_1=0$. Of course, in such a geometrical
configuration one of the geometrical parameter, e.g., either $r_n$ or $z_n$ has to be chosen arbitrarily, 
and the other
one can be expressed in terms of the known one. Further, to choose one of the parameters, we assume that
the nucleus is incompressible, as a consequence, the density will not change even if the geometrical
configuration has changed. Then we have the simple relation $r_n^2z_n=3A/4r_0^3$. In this expression, if
we choose one of the unknown arbitrarily (with the numerical value very close to the nuclear radius), the 
other can also be known. However, we have noticed that in this case
the degree of degeneracy is even more than the spherically symmetric case. Along the curved
surface for the nuclei, for $r_1=r_n$, we can have along the positive direction of $z$-axis, any value of 
$z$ from $0$ to $z_n$. Whereas on the plane faces, we have $z=z_n$ and $r_1$ can have any value from $0$ to $r_n$.
All these points are on the cylindrically deformed nuclear surface and the potential on the surface 
is given by eqn.(52). Hence we may conclude that a non-degenerate solution can only be obtained, if and
only if the electron distribution is spherically symmetric and the nucleus is also spherical with a common
centre (concentric spheres). Which we have studied previously \cite{R5}. However, we expect that if both the 
electron distribution and the shape of the nucleus at the centre are
ellipsoidal in nature with the major axis (which is common for both the ellipsoids) along the magnetic field 
and two other axes (again common for both of them) are symmetric, the
problem of degenerate solutions for electric potential inside WS cells will be removed. At present we are 
persuing this analysis. 
We believe that with this type of geometrical configuration, the electric field at the surface of the WS cell 
will also vanish, which is non-zero in the cylindrical case as discussed below..  

Although the WS cells are overall charge neutral, at any point $(r_{max},z)$ on the curved surface and 
at any point $(r,z_{max})$ on the plane
faces the potential $\psi(r,z)$ can not be a constant, here $r_{max}$ and $z_{max}$ are transverse and
half longitudinal dimensions. Which further means that the electric field at
the surface (both radial and longitudinal components) can not be zero. This is a purely geometrical
effect and such a deformed charge distribution exhibit quadrupole moment.
The modified form of electro-static potential at any point on the curved face of the cylindrically
deformed WS cell is given by 
\begin{equation}
\psi(s,\theta)=CJ_0(\xi r_{max})\exp({-\Lambda s \sin \theta}),
\end{equation}
whereas on the plane faces, this potential is given by 
\begin{equation}
\psi(s,\theta)=CJ_0(\xi s \cos \theta)\exp({-\Lambda s \sin \theta})
\end{equation}
where $s=(r_{max}^2+z^2)^{1/2}$ for the curved face and $=(r^2+z_{max}^2)^{1/2}$ for the plane faces,
and $\Lambda=(\lambda^2+\xi^2)^{1/2}$ is a constant. Here the variable $\theta$ is introduced to obtain
the variation of $z$ on the curved surface and also $r$ on the plane faces. Therefore this $\theta$
variable is not the conventional $\theta$ coordinate used in cylindrical system.
From eqns.(18) and (19), it is obvious that the potential can not be constant on the surfaces. The
corresponding electric field on the curved surface is given by 
\begin{eqnarray}
\vec E&=&\frac{\partial \psi}{\partial s}\hat{e_s}+\frac{1}{s}\frac{\partial \psi}{\partial
\theta}\hat{e_\theta}\\
&=&-2C\exp({-\Lambda (s^2-r_{max}^2)^{1/2}})(J_0(\xi r_{max})\Lambda \hat{e_z}+J_1(\xi r_{max})\xi
\hat{e_r})
\end{eqnarray}
Similarly the electric field at the plane faces is given by  
\begin{eqnarray}
\vec E&=&\frac{\partial \psi}{\partial s}\hat{e_s}+\frac{1}{s}\frac{\partial \psi}{\partial
\theta}\hat{e_\theta}\\
&=&-2C\exp({-\Lambda z_{max}})(J_0(\xi (s^2-z_{max}^2)^{1/2})\Lambda \hat{e_z}+J_1(\xi
(s^2-z_{max}^2)^{1/2})\xi \hat{e_r})
\end{eqnarray}
Which are obviously non vanishing on the surface of WS cells. However, neighboring cylindrical WS
cells will interact electro-magnetically because of the non-zero values of electrostatic fields at the
surfaces. Due to some kind of
electro-magnetic induction there will be charge polarization on the surfaces of cylindrically deformed WS
cells. As a result a number of charged (induced by the nearest neighbor WS cells) WS cells will form
a bundle of charge neutral WS cylinders, instead of a single cylindrical cell \cite{RU1}. Again we expect
that with ellipsoidal coordinate system, the non-zero electric field problem at the WS cell surface will
also be solved.

Now from the Thomas-Fermi condition, we have
\begin{equation}
p_F=[\psi^2(r,z)-m_{\nu}^2]^{1/2}
\end{equation}
where $m_{\nu}=(m_e^2+2\nu eB)^{1/2}$.
With this exact expression for $p_F$, the number density for electron gas can be expressed as 
\begin{equation}
n_e(r,z)=\frac{eB}{2\pi^2}\sum_{\nu=0}^{\nu_{max}}(2-\delta_{\nu 0})[\psi^2(r,z)-m_{\nu}^2]^{1/2}
\end{equation}
This is obviously more exact than eqn.(3), where the value of Fermi momentum $p_F$ is taken from eqn.(6). 
Further, this expression shows that the electron density
is a function of both $r$ and $z$ within the WS cells. Which actually justifies the assumption 
that the electron distribution inside each WS cell around the fixed nucleus is non-uniform.
Now from the non-negative nature of $p_F ^2$, we have
\begin{equation}
\nu_{max}=\frac{\psi^2(r,z)-m_e ^2}{2eB}=\nu_{max}(r,z)
\end{equation}
The upper limit of Landau quantum number $\nu_{max}$ will therefore also depend on the positional 
coordinates $r$ and $z$ within and on the curved surface and the plane faces of the cylinder.
In our model, the upper limit of Landau quantum number is zero at the deformed WS cell surface. Had
the electron distribution at the crustal region been homogeneous, like electron gas in a highly conducting
metal, with a background of positively charged nuclei at rest, the value of $\nu_{max}$ would have remain 
same at all points and for all the electrons depending
on the density of electron gas and the strength of magnetic field. However, in our model, the electrons
are distributed in an axially symmetric cylinder around a symmetrically placed positively charged nuclei
of spherical in shape. The electrostatic  potential changes with $r$ and $z$ accordingly following
eqns.(9) and (12). From the nature of such variation, we found that $\nu_{max}$ is maximum near the
nuclear surface and vanishes at the WS surfaces (both curved surface and the plane faces). Therefore
replacing both $r$ and $z$ by their maximum values 
$r_{max}$ and $z_{max}$ respectively, we have 
\begin{equation}
\nu_{max}(r_{max},z_{max})=0
\end{equation}
Further, from the overall charge neutrality of the WS cells we can write 
\begin{equation}
Z=2\pi e \int_{r_n}^{r_{max}} \int_{r_n}^{z_{max}} n_e(r,z)r dr dz=~~{\rm{constant}}
\end{equation}
These two equations (eqns.(28) and (28)) are solved numerically to obtain $r_{max}$ and $z_{max}$ 
for various values of magnetic field strength for the inner crust matter with metallic iron only. 
Knowing both $r_{max}$ and $z_{max}$, if we solve the equations $\nu_{max}(r,z_{max})=0$ and
$\nu_{max}(r_{max},z)=0$ numerically for $r$ (ranges from $0$ to $r_{max}$) and $z$ (ranges from $0$ to 
$z_{max}$), which 
are on the surface of the cylinder, one can generate the whole surface of the cylinder, including the plane
faces. 

In fig.(3) we have plotted electron number density obtained from eqn.(26), in terms of normal nuclear 
density multiplied by $10^4$, as a function of
radial distance from nuclear surface to the WS cell boundary for the magnetic field strengths $10^1$,
$5\times 10^2$, $5\times 10^3$, $10^4$ and $5\times 10^4$ times $B_c^{(e)}$, indicated by the curves 
$a$, $b$, $c$, $d$ and
$e$ respectively (In all the plots, we have used the values of $C$ and $\xi$ for the boundary condition on
the equator and poles of the spherically symmetric nucleus. Although theoretically speaking the
electrostatic potential inside the WS cells are degenerate with respect to the boundary condition on the
nuclear surface, in practice, we have noticed from the numerical calculation that the variation with other
points as the boundary is not so appreciable). These curves show that the electron number density is maximum 
near the nuclear
surface and minimum near the WS cell boundary $r_{max}$. This figure also shows that electron
number density increases with the strength of magnetic field. In fig.(4) we have plotted the same
quantity as in fig.(3) but against the axial distance $z$ from the nuclear surface to the WS cell
boundary (plane faces) indicated by $z_{max}$.
In this case also the variations are exactly same, qualitatively and quantitatively, as in fig.(3). The
little qualitative difference is because
of different types of functional dependencies.
In fig.(5) we have plotted the upper limit of Landau quantum
number $\nu_{max}$ as a function of radial coordinate from the nuclear surface to the WS cell 
boundary. In this figure the upper curve is for $10\times B_c^{(e)}$, 
the middle one is for $10^2\times B_c^{(e)}$ and the lower one is for 
$500\times B_c^{(e)}$. The value of $\nu_{max}$ decreases with the increase in magnetic field strengths.
It has been observed that beyond $500 \times B_c^{(e)}$, $\nu_{max}$ becomes identically zero throughout 
the WS cell. Further, the value of $\nu_{max}$ for $B\leq 500 \times B_c^{(e)}$ is largest near the nuclear
surface and exactly zero near the cell boundary. In fig.(6) we have plotted the same kind of variation
of $\nu_{max}$, but against the axial distance from nuclear surface to the cell boundary. The
qualitative and the quantitative variations are exactly identical with fig.(5). These two figures show that
the electrons are completely spin polarized in the direction opposite to the external magnetic field $\vec B$
at the cell boundary, including the plane faces, even for $B\leq 500 \times B_c^{(e)}$, but beyond this value 
they are polarized at all the points within the cells.
Although the variations along radial and axial directions are shown in these two figures, we expect
that such polarized picture of electron gas exists throughout the cylindrically deformed
WS cell surface, including the two plane faces. To explain the phenomenon of electron
spin polarization within the WS cells, one has to solve Dirac equation in presence
of strong magnetic field. It is then trivial to show that the Landau quantum number $\nu$ is given by: $2\nu
=2n+1+\alpha$, where $n=0,1,2,.....$ is the principal quantum number and $\alpha=\pm 1$ are the
eigen-values of spin operator $\sigma_z$ corresponding to up and down spin states of electrons
respectively. Hence it is quite obvious that all the Landau levels with $\nu \neq 0$ are doubly
degenerate (with two possible combinations of $n$ and $\alpha$), whereas zeroth order Landau level is 
singly degenerate with only possible combination is
$\nu=0$ is $n=0$ and $\alpha=-1$. Which actually mean that in the zeroth Landau level spins of all the
electrons are aligned in the direction opposite to the external magnetic field. This is a kind of
electron spin polarization under the influence of ultra-strong magnetic field, occurring at the
surface region of WS cells, and also throughout the cells beyond some magnetic field strength. This type of 
spin polarization will not be observed if electrons satisfy schr${\ddot{\rm{o}}}$dinger equation, even
if we introduce electron spin by hand. 
Therefore, the spin polarization is a purely relativistic effect.
In fig.(7) we have shown the
variation of $z_{max}$ with the strength of magnetic field $B$. This figure shows that the variation is
almost insensitive for low and moderate magnetic field strengths but decreases almost abruptly beyond
$10^{16}$G, when most of the electrons occupy their zeroth Landau level, and finally 
tends to saturate to a constant value $\sim 10$fm. The variation
of $r_{max}$ with the strength of magnetic field is shown in fig.(8). The nature of variation is more
or less same as
that of $z_{max}$. However, we have noticed that for extremely large field strength, 
$r_{max}\longrightarrow 0$, instead of saturation.
This is a remarkable difference from its longitudinal counter part. It actually shows that in presence
of extremely strong magnetic field the cylindrically deformed WS cells become more and more thin in the 
transverse direction. We therefore conclude that with the increase in magnetic field
strength the radial contraction will be enormous compared to the axial one. From figs.(7)-(8) we have noticed
that the variations are most significant  beyond $B=10^{16}$G. The reason is again because the electrons 
occupy only their zeroth Landau level in presence of such strong magnetic field, 
at which the quantum mechanical effect of the magnetic field dominates. 
In these two figures we have taken $Z=26$ and $A=56$, which are the atomic  number and mass number 
respectively for the deformed iron atoms. 

Next we calculate the different kinds of energies associated with the electron gas within WS cells. 
The cell averaged kinetic energy density of the electron gas is given by
\begin{eqnarray}
\epsilon_k=\frac{2}{V}\frac{eB}{2\pi^2}\int d^3r&&\sum_{\nu=0}^{\nu_{max}} (2-\delta_{\nu 0}) \nonumber \\ &&
\int_0^{p_F(r,z)} dp_z [(p_z^2+m_\nu^2)^{1/2}-m_e]
\end{eqnarray}
where in the cylindrical coordinate system with azimuthal symmetry $d^3r=2\pi rdr dz$, with the limits
$r_n \leq r \leq r_{max}$ and $r_n \leq z \leq z_{max}$ and $V=\pi r_{max}^2 2z_{max}$, the
volume of each cell. 
The $p_z$ integral is trivial, which gives an analytical expression for the local kinetic energy
density i.e., at a particular point $(r,z)$ within the WS cell. The factor $2$ in the expression for $V$ is 
for $z$-symmetry about $z=0$ plane. 
In fig.(9) we have plotted the variation of local kinetic energy density as a function of radial distance $r$ 
from the nuclear surface to the cell boundary for various magnetic field values, keeping $z=0$. 
This figure shows that  the kinetic energy density increases with the
increase in magnetic field strength. We have indicated the curves by $a$, $b$, $c$, $d$ and $e$ for
the magnetic field strengths $10B_c^{(e)}$, $5\times 10^2B_c^{(e)}$, $5\times 10^3B_c^{(e)}$, 
$10^4B_c^{(e)}$ and
$5\times 10^4B_c^{(e)}$ respectively. In fig.(10) we have shown the same kind of variations along axial
distance from the nuclear surface to one of the plane faces of the WS cell. The variation with
magnetic field strength is again almost identical with fig.(9). Similar to the variation of 
electron number density within the cell (fig.(3) and fig.(4)), both from fig.(9) and fig.(10) 
it may be concluded that the local kinetic energy density for electron gas is maximum 
near the nuclear surface and minimum at the cell boundary. 

Similarly, the cell averaged electron-nucleus interaction energy per unit volume is given by 
\begin{equation}
E_{en}=-\frac{2}{V} Ze^2 \int d^3r \frac{n_e(r,z)}{(r^2+z^2)^{1/2} }
\end{equation}
Analogous to the local kinetic energy density, here also one can obtain the local interaction energy per unit
volume, $E_{en}(r,z)$ at a particular point within the cell. In fig.(11), the variation of the magnitude 
of electron-nucleus interaction energy per unit volume with the radial distance is plotted. 
In this figure 
we have indicated the curves by $a$, $b$, $c$, $d$ and $e$ for
the magnetic field strengths $10B_c^{(e)}$, $5\times 10^2B_c^{(e)}$, $5\times 10^3B_c^{(e)}$, 
$10^4B_c^{(e)}$ and $5\times 10^4B_c^{(e)}$ respectively. This figure shows that the magnitude 
of electron-nucleus local interaction energy increases with the increase in magnetic field strength. 
Which means that with the increase in magnetic field strength the electrons become more strongly 
bound by the Lorentz force of the form $e\vec v\times\vec B$. 
In fig.(12) we have shown the same kind of variation along z-axis. Both the qualitative and the
quantitative nature of variations with the strength of magnetic field are same as that of fig.(11). 

It is also obvious from figs.(9)-(12) that for a given magnetic field
strength the kinetic energy density and the magnitude of electron-nucleus interaction energy per unit
volume at a particular point, either along axial direction or in the radial direction, inside WS
cells are of the same order of magnitude. We do expect that the same type of variations will be obtained
at all the points inside the WS cells. 
In fig.(13) we have shown the variation of cell averaged kinetic energy of
electron gas (solid curve) and the magnitude of cell averaged electron-nucleus interaction energy 
(dashed curve) with the magnetic field strength. This figure shows that both the quantities are increasing 
function of magnetic field strength. However, for low and moderate field values, the variations are not 
very much sensitive, particularly for the kinetic energy part of electron gas.

Next we consider the cell averaged electron-electron direct interaction energy density, given by
\begin{equation}
E_{ee}^{dir}=\frac{1}{V}e^2\int d^3r n_e(r,z)\int d^3r^\prime n_e(r^\prime,z^\prime) \frac{1}{[(\vec r
-\vec r^\prime)^2 +(z-z^\prime)^2]^{1/2}}
\end{equation}
Assuming $\vec r$ as the reference axis, we have $(\vec r-\vec r^\prime)^2 =r^2+{r^\prime}^2
-2rr^\prime\cos \theta$, where $\theta$ is the angle between $\vec r$ and $\vec r^\prime$, assumed to
be on the same plane. Then
$d^3r^\prime=r^\prime dr^\prime d\theta dz^\prime$, with $0\leq \theta \leq 2\pi$. 
In this case the $z$ integral has to be broken into two parts, one with limit $-z_{max} \leq z \leq
\overline z$ and the other one with the limit $\overline z \leq z \leq +z_{max}$. The value of
$\overline z$ is not easy to evaluate in the region between $z$-axis and $r$-axis. With $\theta$
symmetry, for the sake of simplicity we put $\overline z=r_n$ and expect that
the error will be nominal. Now to obtain electron-electron
direct interaction energy one has to evaluate the five dimensional integral as shown in eqn.(32). 
None of them can be obtained analytically, hence it is necessary to follow some numerical methods. 
Even the $\theta$ integral can not be obtained analytically. One can express the $\theta$ integral 
in the form of an elliptical integral of first kind, given by
\begin{equation}
I_{\cal{EL}}(r,r^\prime,z,z^\prime)=\int_0^{\pi/2} d\theta \frac{1}{(1-K\cos^2\theta)^{1/2}}
\end{equation}
where $K=4rr^\prime/[(r+r^\prime)^2+(z-z^\prime)^2]$. The direct part is then given by
\begin{equation}
E_{ee}^{dir}=\frac{4}{2V}e^2\int d^3r n_e(r,z)\int r^\prime dr^\prime dz n_e(r^\prime,z^\prime) \frac{1}{[(r
+ r^\prime)^2 +(z-z^\prime)^2]^{1/2}} I_{\cal{EL}}(r,r^\prime,z,z^\prime)
\end{equation}
where the factor $4$ is coming from the angular integral over $\theta$ from $0$ to $2\pi$.
Let us now consider the
elliptic integral (eqn.(33)) on $z=0$-plane. In this case both $z$ and $z^\prime$ are zero and the
factor $K=1$. Then it can be shown very easily that the integral given by eqn.(33) will diverge 
at the lower limit. To avoid this unphysical infinity we put a lower cut-off (infrared cut-off) $\delta$, 
which will now the lower limit for $\theta$. 
The physical meaning of non-zero lower limit for the $\theta$-integral is that
the two electrons under consideration can not be at zero distance from
each other on the arc of a circle whose centre is same as that of the nucleus. The infrared cut-off
$\delta$ which is a measure of minimum angular distance between two neighboring electrons 
must necessarily depends on the minimum possible linear distance between them and also on the average
radial distance from the centre of the nucleus. From a very elementary geometrical construction 
it can be shown that 
\begin{equation}
\delta=\frac{s}{r}
\end{equation}
where $s$ is the arc length, or the distance between two neighboring electrons on the circular arc. 
Since $s$ is infinitesimal in nature, we can approximate it by a straight
line of length $s$, which is the length of the cord connecting two points occupied by two neighboring
electrons. Since $s$ is the minimum possible linear distance between two electrons, we can express
it in terms of electron density near those points, given by 
\begin{equation}
s\sim n_e^{-1/3}
\end{equation}
The cell averaged electron-electron direct energy has been obtained by evaluating the
multi-dimensional integrals numerically.

The electron-electron exchange energy corresponding to the $i$th electron in the cell is given by
\begin{equation}
E_{ee}^{(ex)}= -\frac{e^2}{2}\sum_j \int d^3r d^3r^\prime \frac{1}{ [(\vec
r-\vec {r^\prime})^2+(z-z^\prime)^2]^{1/2}} \bar\psi_i(\vec r,z)\bar\psi_j(\vec {r^\prime},z^\prime)
\psi_j(\vec r,z)\psi_i(\vec {r^\prime},z^\prime)
\end{equation}
where $\psi_i(r,z)$ is the spinor wave function of Dirac equation in cylindrical coordinate in presence of 
strong quantizing magnetic field, and $\bar \psi({\vec r},z)=
\psi^\dagger({\vec r},z)\gamma_0$, the adjoint of the spinor and $\gamma_0$ is
the zeroth part of the Dirac gamma matrices $\gamma_\mu$ in cylindrical coordinate system.
We have evaluated the cell averaged exchange energy using Dirac spinors in cylindrical coordinate.
An elaborate discussion is given in the Appendix.

The kinetic pressure of non-uniform electron gas within the WS cell is given by 
\begin{equation}
P(r,z)=\frac{eB}{2\pi^2}\sum_{\nu=0}^{\nu_{max}(r,z)}(2-\delta_{\nu 0})\int_0^{p_F(r,z)}
\frac{p_z^2dp_z}{(p_z^2+m_\nu^2)^{1/2}} 
\end{equation}
The $p_z$ integral is very easy to evaluate analytically and is given by 
\begin{eqnarray}
P(r,z)&=&\frac{eB}{4\pi^2}\sum_{\nu=0}^{\nu_{max}(r.z)}\big [p_F(p_F^2+m_\nu^2)^{1/2}\nonumber \\ &-& m_\nu^2
\ln\left \{ \frac{p_F+(p_F^2+m\nu^2)^{1/2}}{m_\nu}\right \}\big ]
\end{eqnarray}
which gives the local pressure at $(r,z)$. This equation shows that the electron kinetic pressure 
also changes from point to point within the WS cells. 
In fig.(14) we have shown the variation of kinetic pressure for the non-uniform electron gas with the
radial distance within the cell. Curves $a$ and $b$ are for $B=10\times B_c^{(e)}$ and 
$B=500\times B_c^{(e)}$, whereas upper curve and lower curve as indicated by $c$ (almost identical) are for   
$B=5000\times B_c^{(e)}$ and $B=10000\times B_c^{(e)}$ respectively. In fig.(15) the same kind of
variations are shown along z-axis. In this figure the curves for 
$B=5000\times B_c^{(e)}$ and $B=10000\times B_c^{(e)}$ are almost identical and indicated by single thick
curve $c$. For $B=10\times B_c^{(e)}$ and 
$B=500\times B_c^{(e)}$ the curves are indicated by $a$ and $b$ respectively. These two figures show
that the kinetic pressure is maximum near the nuclear surface and zero at the cell boundary. The
variation with magnetic field strength shows that the non-uniform electron gas becomes softer for
higher magnetic field within the cells. In fig.(16) we have shown the variation of cell averaged 
electron-electron direct
interaction energy (solid curve) and the corresponding kinetic pressure (dashed curve) with
the magnetic field strength. Both the quantities are monotonically increasing function of magnetic field
strength. Since in TOV equation for neutron stars, in the equation of state the cell averaged electron gas 
kinetic pressure is used, hence we can conclude that since electron gas becomes harder in presence of 
strong magnetic field in this particular region, the width of inner crust region will increase with
the increase in magnetic field strength. In fig.(17) we have shown the variation of cell averaged 
electron-electron exchange interaction energy, assuming the extreme case when electrons occupy only their
zeroth Landau level (dashed curve) and the most general one, when electrons can have all possible
Landau levels (solid curve). The figure shows that for $\nu\neq 0$, the exchange energy is oscillatory
for low and moderate values of magnetic field strength. This oscillatory phenomenon is identical with the 
observed De Haas-van Alphen oscillation observed in Landau diamagnetism. The exchange energy becomes extremely 
small 
when the electron Fermi momentum suddenly becomes zero for some value of magnetic field strength and then rises 
sharply upto a certain magnitude of magnetic field
strength and finally decreases to zero, when the magnetic energy dominates over the matter part. In
the case of $\nu=0$, the nature of variation is almost identical, except the oscillatory nature at low
and moderate magnetic field region.

For the sake of completeness, we have obtained the classical form of electron-electron Coulomb potential
energy and the corresponding electron-nucleus interaction energy within a WS cell. The electron-electron 
interaction energy is given by (an elaborate discussion is given in the Appendix)
\begin{eqnarray}
E_{ee}&=&\frac{Z^2 e^2}{r_{max}^4 z_{max}^2}\int_0^\infty \frac{dk}{k^5} [r_{max}J_1(k r_{max}) - r_nJ_1(k
r_n)]^2 \nonumber \\ && [k(z_{max}-r_n) - (1-\exp \{-k(z_{max} - r_n)\})]
\end{eqnarray}
and the corresponding electron-nucleus interaction part is given by
\begin{equation}
E_{en}=-\frac{Z^2 e^2}{r_{max}^2 z_{max}}I
\end{equation}
where 
\begin{eqnarray}
I&=&\frac{1}{2}\left[(z_{max}^2 + r_{max}^2)^{1/2}z_{max} + r_{max}^2 \log \left(\frac{(z_{max}^2 + r_{max}^2)^{1/2} +
z_{max}}{r_{max}}\right)\right] \nonumber \\ 
&-&\frac{1}{2}\left[(r_n^2 + r_{max}^2)^{1/2}r_n + r_{max}^2 \log \left(\frac{(r_n^2 + r_{max}^2)^{1/2} +
r_n}{r_{max}}\right)\right] \nonumber \\ 
&-&\frac{1}{2}\left[(z_{max}^2 + r_n^2)^{1/2}z_{max} + r_n^2 \log \left(\frac{(z_{max}^2 + r_n^2)^{1/2} +
z_{max}}{r_n}\right)\right] \nonumber \\ 
&+&\frac{1}{2}[2^{1/2}r_n^2 + r_n^2 \log (2^{1/2}+1)] \nonumber \\ 
\end{eqnarray}
The detail derivations for electron-nucleus Coulomb interaction energy is also given in the Appendix. 
In fig.(18) we have shown the variation of
electron-electron Coulomb energy density (solid curve) and the corresponding magnitude of electron-nucleus 
classical 
interaction energy density (dashed curve) with the strength of magnetic field. 
Both the curves show that the classical
interaction energies are also insensitive to the strength of magnetic field in the low and moderate
regions, and both of them are affected significantly beyond $10^{16}$G. The reason is again that the
electrons occupy only their zero-th Landau level. However, unlike the quantum mechanical cases, 
here at high magnetic field region both of them become extremely small. 
Moreover the overall magnitude does not change by an order of magnitude within the range of 
magnetic field considered. Finally to show that the cylindrically deformed WS cell structure of the
inner crust matter of strongly magnetized neutron star (magnetar) is energetically favorable over the spherical
structure, in fig.(19) we have compared the energy per electron for these two possible type of WS
cell structures for various values of magnetic field strength. In this figure solid curve indicated by
sph is for spherical cell structure and the dashed one indicated by cyl is for the cylindrically
deformed WS cells. The energy per electron plotted along left y-axis is for the spherical case whereas 
for the cylindrical case the same quantity is plotted along right y-axis. This figure shows that for
cylindrical case the total energy per electron is about one order of magnitude less than that of
spherical case. Further beyond $B\sim 10^{17}$G, energy per electron for the cylindrical case becomes
several orders of magnitude less than the spherical case (see also \cite{RU2}). 
\section{Conclusions}
In this article we have investigated various physical properties of non-uniform electron gas 
assuming cylindrically deformed
atoms of metals, in particular the metallic iron at the inner crust of a strongly magnetized neutron star.
Because of extremely strong surface magnetic field of magnetars, we
have assumed a cylindrical type deformation of the atoms, which are subsequently replaced by WS cells 
with the same kind of geometrical structure. 
The longitudinal axis of all the cylinders are along the direction of magnetic lines of forces. 
The curved surfaces of these cylinders
are therefore parallel to the boundary surface of the neutron stars in the region far away from the
magnetic poles, where the magnetic lines of forces emerge almost perpendicularly with the surface
(polar cap). We have studied various physical quantities for the dense electron gas in the inner crust
region. We have investigated the variations with magnetic field strength for cell averaged quantities and 
also the spatial variations for constant magnetic field.
We have noticed that the transverse dimension of a cylindrically deformed
WS cell becomes extremely thin in presence of ultra-strong magnetic field. In this work we have
also compared the total energy per electron with the spherical case and found that the cylindrically
deformed WS structure in the inner crust region is energetically more favorable over the spherical case in
presence of ultra-strong magnetic field. Although we have considered
cylindrical type deformation for the iron atoms in this  region, it is expected that the 
atoms become cigar shape in presence of strong magnetic field. The present investigation, assuming
cylindrically deformed atoms is therefore an approximate model calculation. 
In our future study, the properties of neutron star inner crust matter with cigar
shape atoms in the metallic crystal in presence of strong magnetic field will be investigated. 
\section{Appendix}

\noindent A. Coulomb energy of electron gas:\par
1. Electron-nucleus Coulomb interaction energy:\par
\begin{equation}
E_{en}=-Ze \int_v \frac{dq}{s}
\end{equation}
Where the integral is over the whole volume and $s=(r^2+z^2)^{1/2}$.
The charge element within an elementary volume $dv$ is given by
\begin{equation}
dq=Ze \frac{dv}{v}=Ze \frac{r dr dz}{2r_{max}^2 z_{max}}
\end{equation}
Hence 
\begin{equation}
E_{en}=-\frac{Z^2 e^2}{2 r_{max}^2 z_{max}} \int_{r_n}^{r_{max}} r dr \int_{r_n}^{z_{max}}
\frac{1}{(r^2+z^2)^{1/2}} 
\end{equation}
The double integral can very easily be evaluated and finally we get eqn.(40) as given in the text.\par
2. Electron-electron Coulomb interaction energy:
\begin{equation}
E_{ee}= \frac{1}{2} \int \int \frac{dq_1 dq_2}{s}
\end{equation}
Where $s=[(z_1-z_2)^2 + (\vec r_1-\vec r_2)^2]^{1/2}$.
In this case the charge elements $dq_1$ and $dq_2$ are at $(r_1,\theta_1,z_1)$ and $(r_2,\theta_2,z_2)$
respectively. The factor $1/2$ is to take care of double counting.
From a very simple geometrical construction it can very easily be shown that
\begin{equation}
s=\left[(z_1-z_2)^2 + {(r_1^2+r_2^2-2r_1r_2 \cos (\theta_1-\theta_2))}\right]^{1/2}
\end{equation}
Substituting for $dq_1$ and $dq_2$, with the definition as given in eqn.(42), we have 
\begin{eqnarray}
E_{ee}&=&\frac{1}{2} \frac{Z^2 e^2}{4\pi ^2 r_{max}^4z_{max}^2} \int_{r_n}^{r_{max}}r_1 dr_1
\int_{r_n}^{r_{max}}r_2
dr_2 \int_{r_n}^{z_{max}}dz_1 \int_{r_n}^{z_{max}}dz_2 \int_0^{2\pi} d\theta_1 \int_0^{2\pi} d\theta_2
\nonumber \\ &&
\frac{1}{\left[(z_1-z_2)^2 + {(r_1^2+r_2^2-2r_1r_2 \cos (\theta_1-\theta_2))}\right]^{1/2}}
\end{eqnarray}
To evaluate the integrals, we use the identity
\begin{eqnarray}
\frac{1}{s}&=&\frac{1}{\left[(z_1-z_2)^2 + {(r_1^2+r_2^2-2r_1r_2 \cos
(\theta_1-\theta_2))}\right]^{1/2}}\nonumber \\
&=&\frac{1}{\pi} \sum_{m=-\infty} ^{+\infty} \exp \{im(\theta_1-\theta_2)\} \int_0^\infty J_m(kr_1)J_m(kr_2)
\exp \{-k(z_>-z_<)\} dk
\end{eqnarray}
To obtain electron-electron Coulomb energy, we first evaluate the potential at $(r_1,\theta_1,z_1)$ due
to a charge element $dq (r_2,\theta_2,z_2)$.
This is given by
\begin{equation}
\phi(r_1,\theta_1,z_1)= \int \frac{dq(r_2,\theta_2,z_2)}{s}
\end{equation}
Using the identity as defined above (eqn.(48)) and integrating over $z_2$, within the range $r_n\leq
z_2\leq z_{max}$, we get 
\begin{eqnarray}
\phi(r_1,\theta_1,z_1)&=&\frac{Ze}{2\pi r_{max}^2 z_{max}} \sum_{m=-\infty}^{+\infty} \int_0^{+\infty}
\frac{dk}{k} \int_{r_n}^{r_{max}}r_2 dr_2 \int_0^{2\pi} d\theta_2 J_m(kr_1)J_m(kr_2) \nonumber \\ && (2 - \exp
\{-k(z_1-r_n)\} -
\exp\{-k(z_{max}-z_1)\}) \exp \{im(\theta_1-\theta_2)\}
\end{eqnarray}
Integral over $\theta_2$ gives $2\pi\delta_{m0}$, which means only $m=0$ term of the series in the
above identity (eqn.(48)).
Hence, we have
\begin{eqnarray}
\phi(r_1,\theta_1,z_1)&=& \frac{Ze}{r_{max}^2 z_{max}} \int_0^\infty \frac{dk}{k} \int_{r_n}^{r_{max}} r_2
dr_2
J_0(kr_1) J_0(kr_2)\nonumber \\ && \left(2-\exp \{-k(z_1-r_n)\}-\exp\{-k(z_{max}-z_1)\}\right)
\end{eqnarray}
Further, to evaluate the $r_2$ integral within the range $r_n\leq r_2\leq r_{max}$, we use the relation
\begin{eqnarray}
\int_0^x r dr J_m(r)=\frac{x \Gamma (\frac{m+2}{2})}{\Gamma(m/2)} \sum_{l=0}^\infty \frac{(\mid
m\mid +2l+1)\Gamma(\frac{m}{2}+l)}{\Gamma(\frac{m+4}{2}+l)} J_{\mid m\mid +2l+1}(x)
\end{eqnarray}
Since $m=0$, we have in the denominator of the above expression $\Gamma(m/2)=\infty$. Whereas the 
numerator of the first term of $l$ series in
the above relation (i.e., for $l=0$ term) also contains a $\Gamma(m/2)=\infty$ term. 
These two terms will cancel each other and we get non-zero contribution for $l=0$ only with $m=0$. 
To evaluate $r_2$ integral we decompose it in the following form as given below
\begin{equation}
\int_{r_n}^{r_{max}}.......dr_2=\int_0^{r_{max}}......dr_2 - \int_0^{r_n}.....dr_2
\end{equation}
Then using eqn.(52) with $l=0$ and $m=0$, it is possible to evaluate $r_2$ integral analytically. 
Then we have the Coulomb
potential due to the charge element $dq(r_2,\theta_2,z_2)$ 
\begin{eqnarray}
\phi(r_1,z_1)&=&\frac{Ze}{r_{max}^2 z_{max}}\int_0^\infty \frac{dk}{k^2}\left[r_{max} J_1(kr_{max})-r_n
J_1(kr_n)\right] \nonumber \\ && J_0(kr_1)\left[2-\exp\{-k(z-r_n)\}-\exp\{-k(z_{max}-z_1)\}\right]
\end{eqnarray}
Hence it is also possible to evaluate the components of electric field,
$E_z=-\partial\phi/\partial z_1$
along z-direction and $E_r=-\partial\phi/\partial r_1$
along the radial direction. It is trivial to show that these components are non-zero on the WS cell
surface, including the plane faces.
The Coulomb energy is then given by 
\begin{equation}
E_{ee}=\frac{1}{2} \frac{Ze}{2\pi r_{max}^2 z_{max}}\int \phi(r_1,z_1,\theta_1) dq(r_1,z_1,\theta_1)
\end{equation}
The $\theta_1$ integral will give $2\pi$, the $z_1$ integral is very easy to evaluate and for $r_1$
integral we use the Bessel integral formula as given in eqn.(52).
Then we have the electron-electron Coulomb interaction energy 
\begin{eqnarray}
E_{ee}=\frac{Z^2 e^2}{r_{max}^4 z_{max}^2}\int_0^\infty \frac{dk}{k^5}\left[r_{max} J_1(kr_{max})-r_n
J_1(kr_n)\right]^2 \left[k(z_{max}-r_n)-(1-\exp\{-k(z_{max}-r_n)\})\right]
\end{eqnarray}

\noindent B. Electron-electron exchange energy:

In the cylindrical coordinate system with the external magnetic field along z-axis, which is also the
symmetry axis of the cylinder and considering the gauge $\vec A(\vec r)\equiv B(-y/2,x/2,0)$, we have
the Dirac equation corresponding to upper component $\phi_\lambda(\vec r)$: 
\begin{equation}
\left [E^2-m^2+2\lambda k+\frac{\partial^2}{\partial \rho^2}+\frac{1}{\rho}\frac{\partial} {\partial \rho} +
\frac{\partial^2}{\partial z^2}-k^2 \rho^2\right]\phi_\lambda(\vec r)=0
\end{equation}
where $k=eB/2$ and $\sigma_z \phi(\vec r)=\lambda \phi$ with $\lambda=\pm 1$, eigen-values for spin-up and
spin-down states respectively. Defining $\beta_\lambda=E^2-m^2+2\lambda k$, the above equation can be
written in the following form.
\begin{equation}
\left [\beta_\lambda +\frac{\partial^2}{\partial \rho^2}+\frac{1}{\rho}\frac{\partial}{\partial
\rho}+\frac{\partial^2}{\partial z^2}-k^2\rho^2\right]\phi_\lambda(\rho,z)=0
\end{equation}
writing the solution for the upper component in the separable form: 
$\phi_\lambda(\rho,z)=f_\lambda (\rho)\exp (ip_zz)$, we have from the above equation 
\begin{equation}
\left [\beta_\lambda +\frac{\partial^2}{\partial \rho^2}+\frac{1}{\rho}\frac{\partial}{\partial
\rho}-k^2\rho^2\right ]f_\lambda(\rho)=0
\end{equation}
where $\beta_\lambda$ is replaced by $\beta_\lambda-p_z^2$.
The solutions are given by 
\begin{equation}
f_\lambda(\rho)=\exp\left (-\frac{t}{2}\right )g_\lambda(t)
\end{equation}
where $t=\eta^2=k\rho^2$. Substituting $f_\lambda (\rho)$, we have
\begin{equation}
g_\lambda=\frac{N}{L}^{1/2}\exp(ip_zz)\exp\left (-\frac{t}{2}\right )L_n(t)
\end{equation}
where $L$ is the linear dimension along z-axis, $L_n(t)$ is the well known Laguerre polynomial,  
\begin{equation}
\mid N\mid = \left [\frac{k}{2E}\left (1+\frac{m}{E}\right )\right ]^{1/2}
\end{equation}
is the normalization constant. Considering the effect of spin into account, we have 
\begin{equation}
g_{\lambda=+1}=L_n\left (\frac{1}{2} eB\rho^2\right )\left (\begin{array}{c} 1\\0\end{array} \right )
\end{equation}
and 
\begin{equation}
g_{\lambda=-1}=L_n\left (\frac{1}{2} eB\rho^2\right )\left (\begin{array}{c} 0\\1\end{array} \right )
\end{equation}
are the up-spin and down-spin states respectively for the upper component, with the
corresponding energy eigen-values $E_+=(p_z^2+m^2+2neB)^{1/2}$ and $E_+=(p_z^2+m^2+2(n+1)eB)^{1/2}$.

The spin-up and spin-down spinor states are then given by 
\begin{equation}
\psi^\uparrow (\rho,z)=\left [\frac{k}{2E}\left (1+\frac{m}{E}\right )\right
]^{1/2}\frac{\exp(ip_zz)\exp\left (-\frac{t}{2}\right )}{{L}^{1/2}}\left (\begin{array}{c} L_n\\ 0\\ p_z
L_n/(E+m)\\ -2ik^{1/2}t^{1/2}/(E+m) L_n^\prime \end{array}\right)
\end{equation}
and
\begin{equation}
\psi^\downarrow (\rho,z)=\left [\frac{k}{2E}\left (1+\frac{m}{E}\right )\right
]^{1/2}\frac{\exp(ip_zz)\exp\left (-\frac{t}{2}\right )}{{L}^{1/2}}\left (\begin{array}{c} 0\\ L_n\\
-2ik^{1/2}t^{1/2}/(E+m)(L_n^\prime-L_n)\\ -p_z L_n/(E+m) \end{array}\right)
\end{equation}
Substituting for up-spin state, we have 
\begin{eqnarray}
\overline\psi_i^\uparrow(\vec r)\psi_i^\uparrow(\vec r^\prime)&=&\frac{k}{2\pi L}\left
(1+\frac{m}{E}\right)\exp\{-ip_z(z-z^\prime)-\frac{1}{2}(t+t^\prime)\}\nonumber \\ && \left \{\left
(1-\frac{p_z^2}{(E+m)^2}\right)L_n(t)L_n(t^\prime) -
\frac{4k(tt^\prime)^{1/2}}{(E+m)^2}L_n^\prime(t)L_n^\prime(t^\prime)\right \}
\end{eqnarray}
Similarly,
\begin{eqnarray}
\overline\psi_i^\downarrow(\vec r^\prime)\psi_i^\downarrow(\vec r)&=&\frac{k}{2\pi L}\left
(1+\frac{m}{E}\right)\exp\left \{-ip_z(z-z^\prime)-\frac{1}{2}(t+t^\prime)\right \}\nonumber \\ && \left [\left
(1-\frac{p_z^2}{(E+m)^2}\right)L_n(t)L_n(t^\prime) -
\frac{4k(tt^\prime)^{1/2}}{(E+m)^2}
\left\{(L_n^\prime(t^\prime)-L_n(t^\prime))(L_n^\prime(t)-L_n(t))\right\} \right ]
\end{eqnarray}
Identical expressions can also be obtained for $j$th type particles, on which there is a sum. To
obtain the exchange energy, we replace the sum over $j$ by the integral over momentum $p_z$ of $j$th
particle, i.e., $\sum_j\rightarrow L\int dp_z^\prime$. Assuming $\vec r$ as the
reference axis, we can write $\int d^3x d^3x^\prime ...=2\pi \int\rho d\rho dz \rho^\prime
d\rho^\prime dz^\prime d\theta^\prime ...$. Replacing the variables $z\prime$ and $p_z^\prime$ by two
new variables, 
$\overline z$ and $P_z$, where $\overline z=z^\prime-z$ and
$P_z=p_z-p_z^\prime$, we have the $\overline z$ integral 
\begin{equation}
I=\int_{-\infty}^{+\infty}\frac{\exp (i\overline zP_z)}{(\overline z^2+X^2)^{1/2}} d\overline z
=2K_0(XP_z)
\end{equation}
where $X=\mid \vec \rho-\vec \rho^\prime \mid$ and
\begin{equation}
K_0(x)=\int_0^\infty \frac{\cos (xt)}{(t^2+1)^{1/2}}dt
\end{equation}
is the modified Bessel function of order zero. The z-integral simply gives $L$. The integral over
$P_z^\prime$ can be obtained analytically using the following integrals (eqns.(71)-(72)): 
\begin{equation}
I=\int_0^{2p_F}K_0(P_z\mid X \mid)dP_z=\frac{1}{\mid X \mid}\int_0^\alpha K_0(w)dw
\end{equation}

where $\alpha=2p_F\mid X\mid$ and $p_F$ is the electron Fermi momentum. The value of the integral
on the right hand side of the above equation is obtained from standard mathematical hand book
\cite{mhb} and is given by
\begin{eqnarray}
\int_0^x K_0(t)dt&=& -(\gamma+\log x/2)~x \sum_{k=0}^\infty \frac{(x/2)^{2k}}{(k!)^2(2k+1)}
\nonumber \\ &+& x\sum_{k=0}^\infty \frac{(x/2)^{2k}}{(k!)^2(2k+1)^2}+ x\sum_{k=1}^\infty
\frac{(x/2)^{2k}}{(k!)^2(2k+1)}\times \nonumber \\ && (1+1/2+...+1/k) = xI(x)
\end{eqnarray}
where $\gamma$ (Euler's constant) = $0.5772156649$.
The rest three integrals, over $\rho$, $\rho^\prime$ and $\theta$ are evaluated
numerically. The Laguerre polynomials $L_n(x)$ and their derivatives $L_n^\prime(x)$ are obtained
numerically from the recursion relations and derivative formulas from the reference as cited above and finally
obtained the exchange energy as a function of magnetic field strength as defined in eqn.(36).


\begin{figure*}
\resizebox{0.75\textwidth}{!}{%
\includegraphics{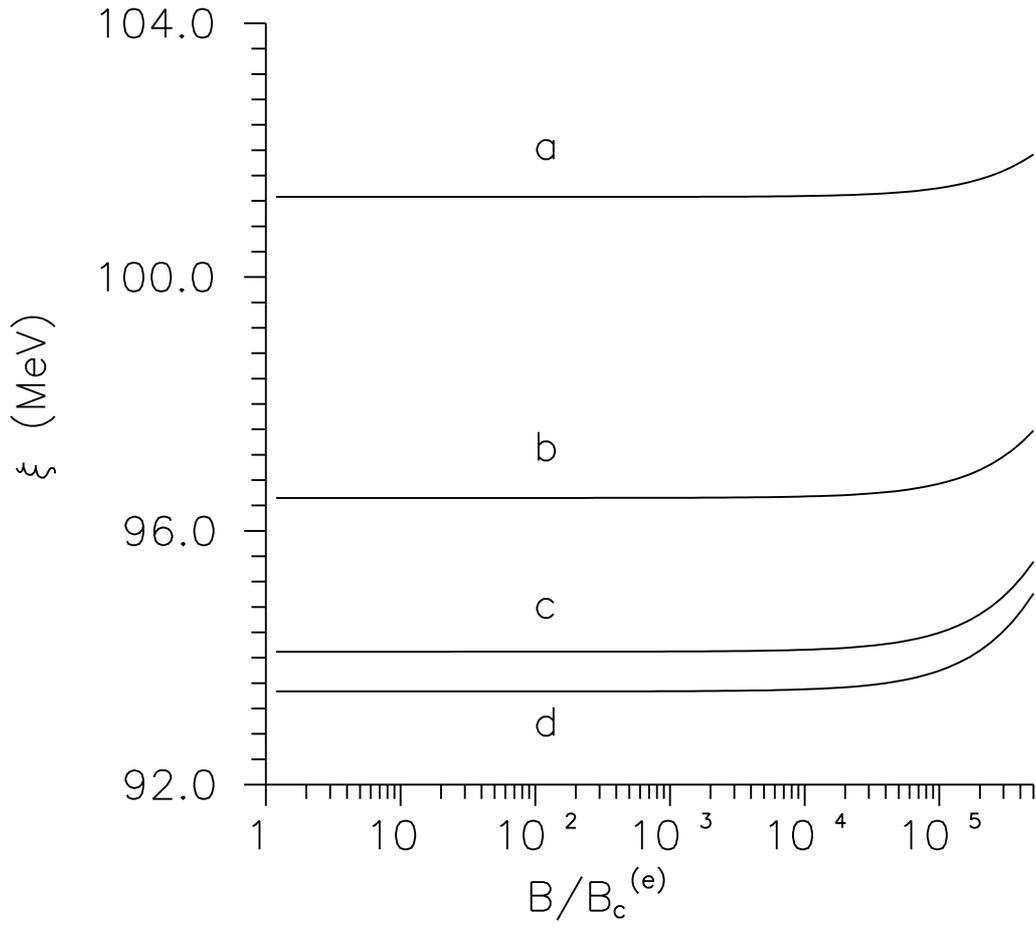}}
\caption{The variation of separation variable $\xi$ with the magnetic field strength. Curve (a): $r$
and $z$ are on the equator and poles respectively, curve (b): $r=r_n/4$, curve (c): $r=r_n/2$
and curve (d): $r=3r_n/4$. For the curves (b),(c) and (d), $z=(r_n^2-r^2)^{1/2}$.}
\label{fig:1} 
\end{figure*}
\begin{figure*}
\resizebox{0.75\textwidth}{!}{%
\includegraphics{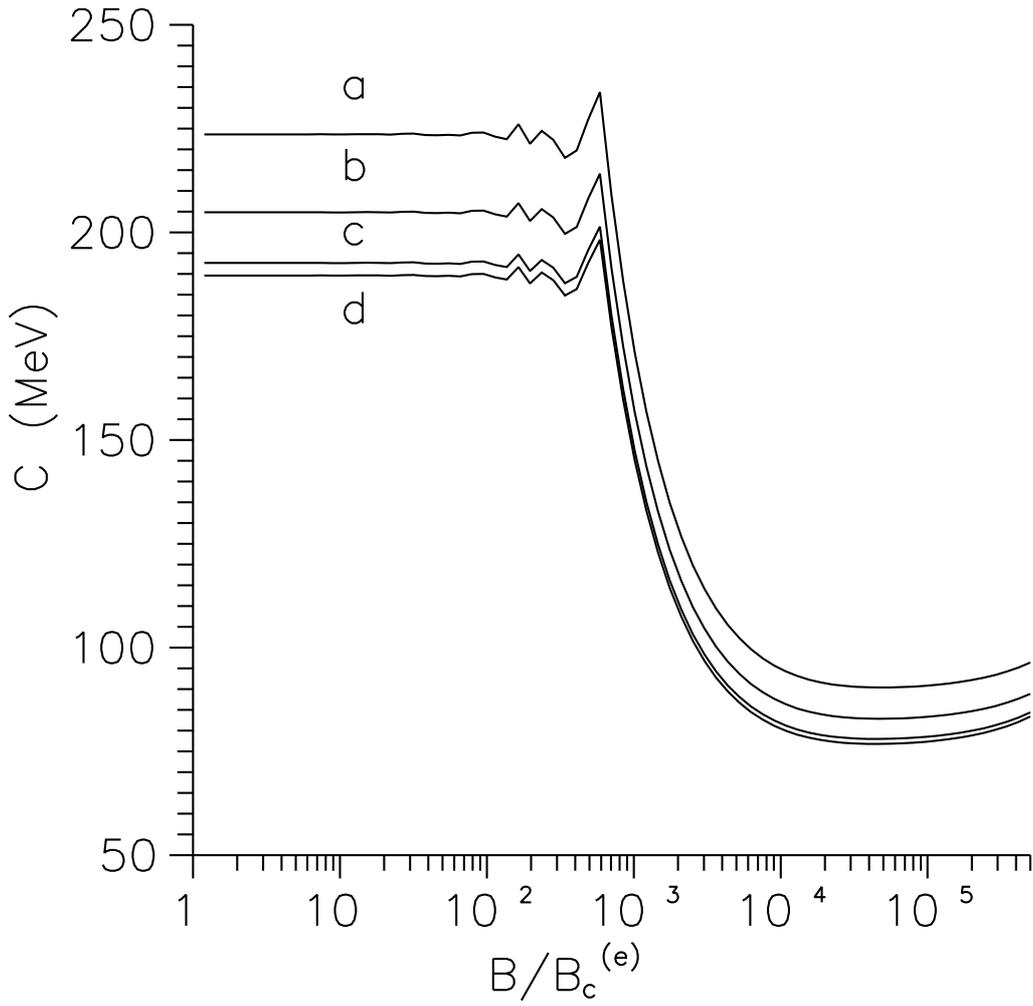}}
\caption{The variation of normalization constant $C$ with the magnetic field strength. Curve (a): $r$
and $z$ are on the equator and poles respectively, curve (b): $r=r_n/4$, curve (c): $r=r_n/2$
and curve (d): $r=3r_n/4$. For the curves (b),(c) and (d), $z=(r_n^2-r^2)^{1/2}$.}
\label{fig:2}
\end{figure*}
\begin{figure*}
\resizebox{0.75\textwidth}{!}{%
\includegraphics{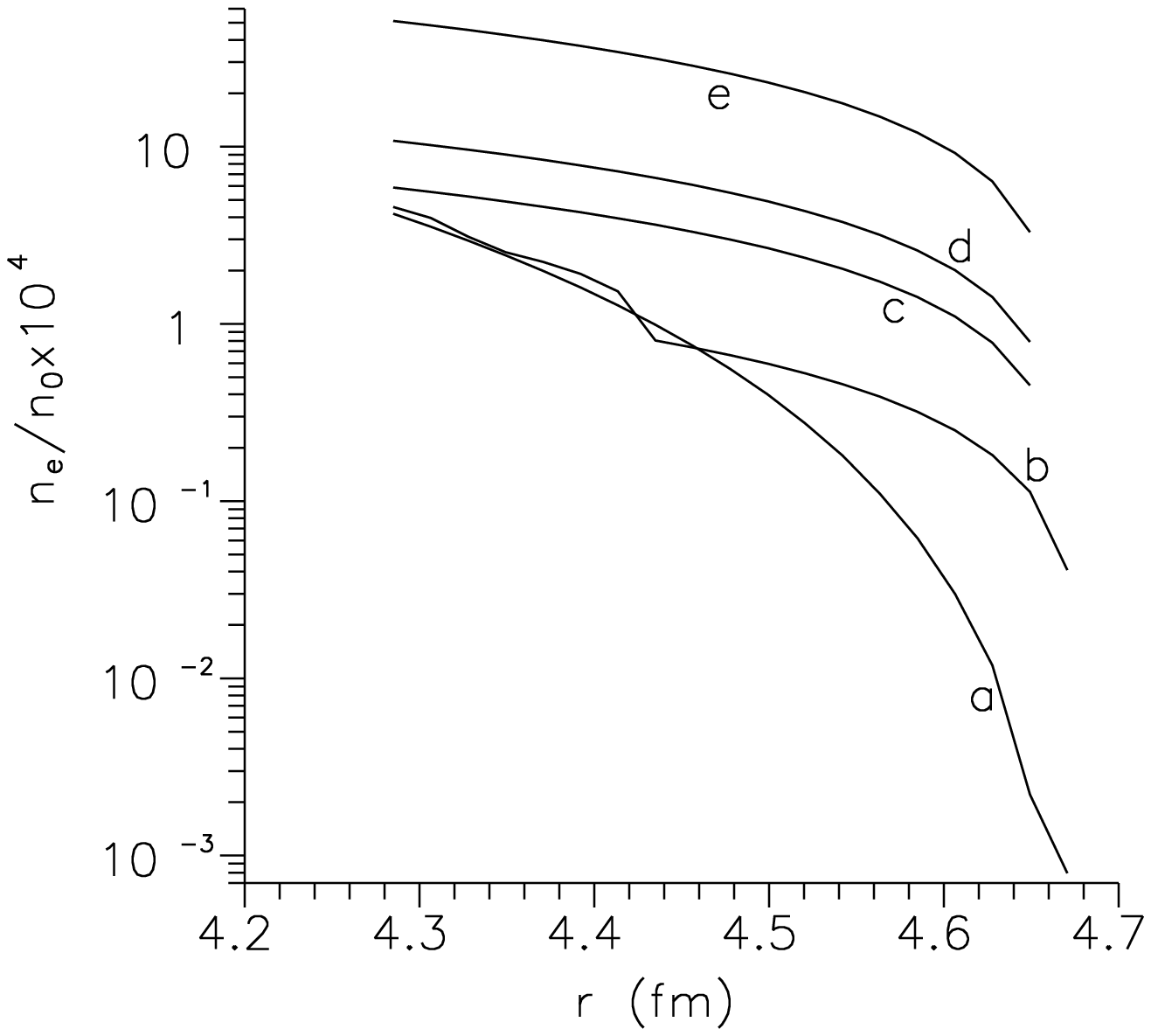}}
\caption{The variation of electron number density expressed in terms of normal nuclear density with
radial distance $r$ in Fermi. The curves indicated by $a$, $b$, $c$, $d$ and $e$ are for $B=10,
5\times 10^2, 5\times 10^3, 1\times 10^4$ and $5\times 10^4$ times $B_c^{(e)}$ respectively.}
\label{fig:3}
\end{figure*}
\begin{figure*}
\resizebox{0.75\textwidth}{!}{%
\includegraphics{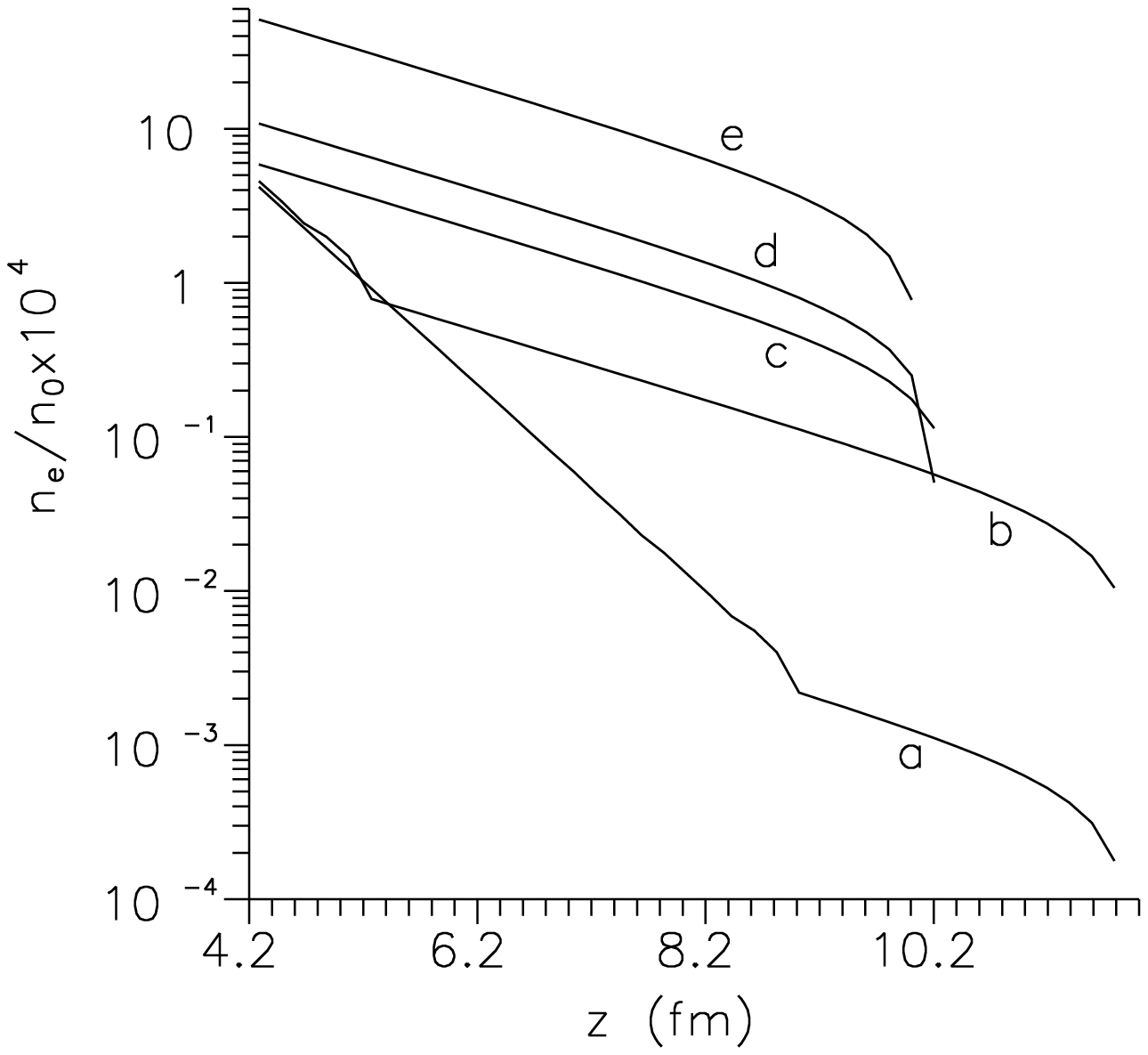}}
\caption{The variation of electron number density expressed in terms of normal nuclear density with
radial distance $z$ in Fermi. The curves indicated by $a$, $b$, $c$, $d$ and $e$ are for $B=10,
5\times 10^2, 5\times 10^3, 1\times 10^4$ and $5\times 10^4$ times $B_c^{(e)}$ respectively.}
\label{fig:4}
\end{figure*}
\begin{figure*}
\resizebox{0.75\textwidth}{!}{%
\includegraphics{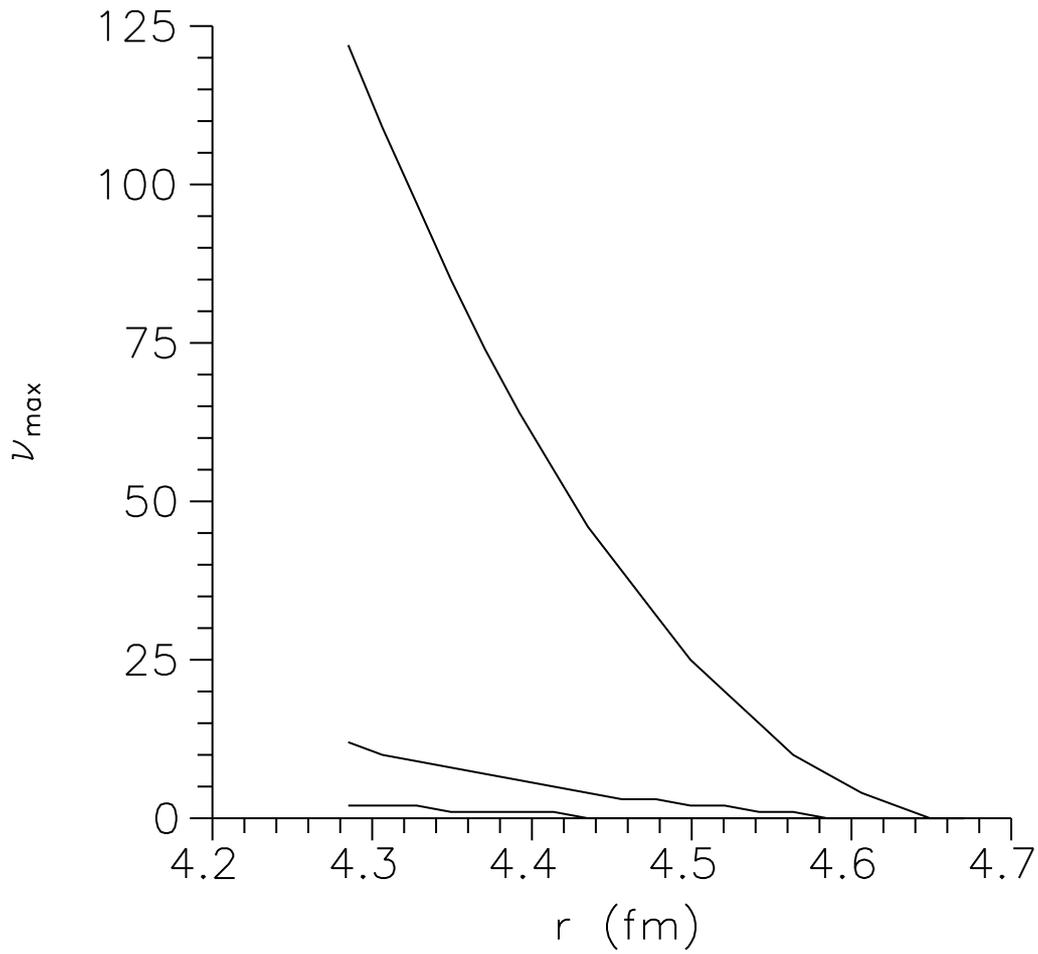}}
\caption{The variation of upper limit of Landau quantum number with the radial distance in Fermi.
Upper curve is for $B=10\times B_c^{(e)}$, middle one is for $10^2\times B_c^{(e)}$ and the lower one
is for $5\times 10^2 B_c^{(e)}$.}
\label{fig:5}
\end{figure*}
\begin{figure*}
\resizebox{0.75\textwidth}{!}{%
\includegraphics{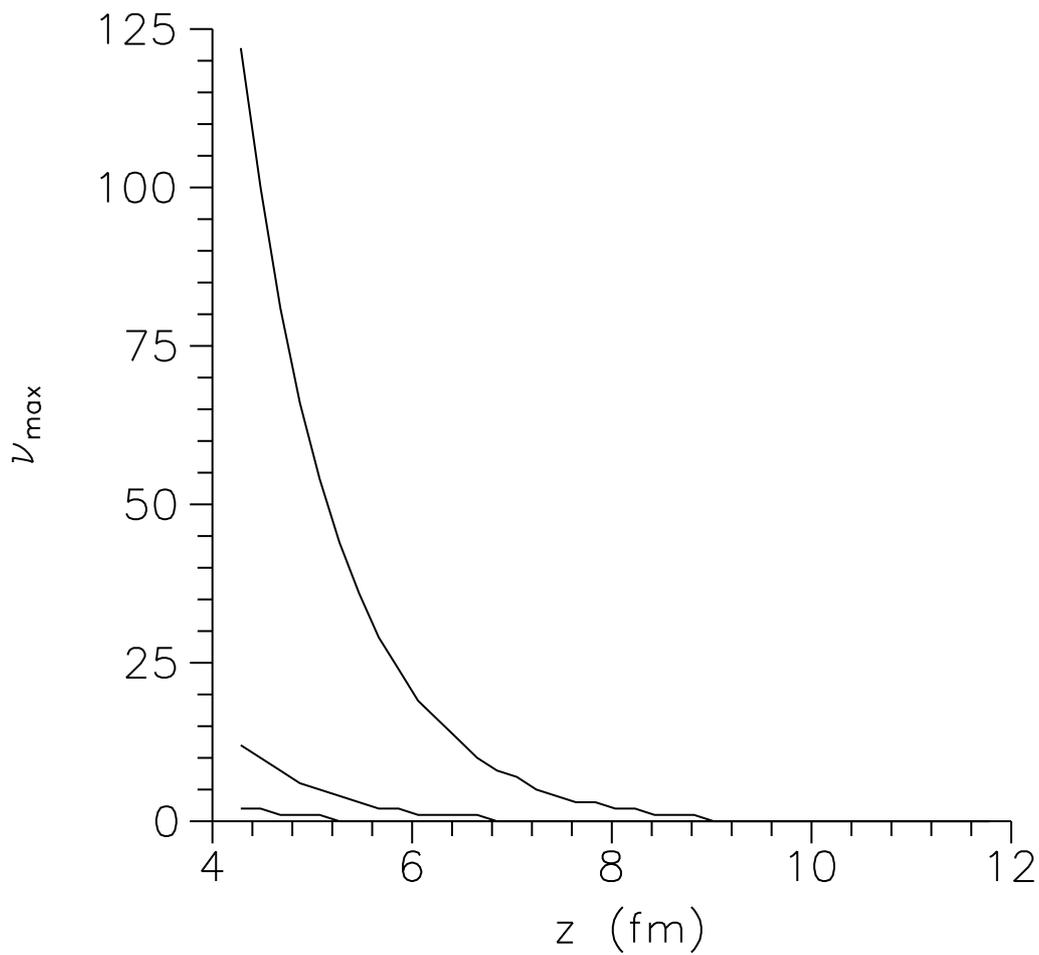}}
\caption{The variation of upper limit of Landau quantum number with the axial coordinate. Upper curve
is for $B=10\times B_c^{(e)}$, middle one is for $10^2\times B_c^{(e)}$ and the lower one
is for $5\times 10^2 B_c^{(e)}$.}
\label{fig:6}
\end{figure*}
\begin{figure*}
\resizebox{0.75\textwidth}{!}{%
\includegraphics{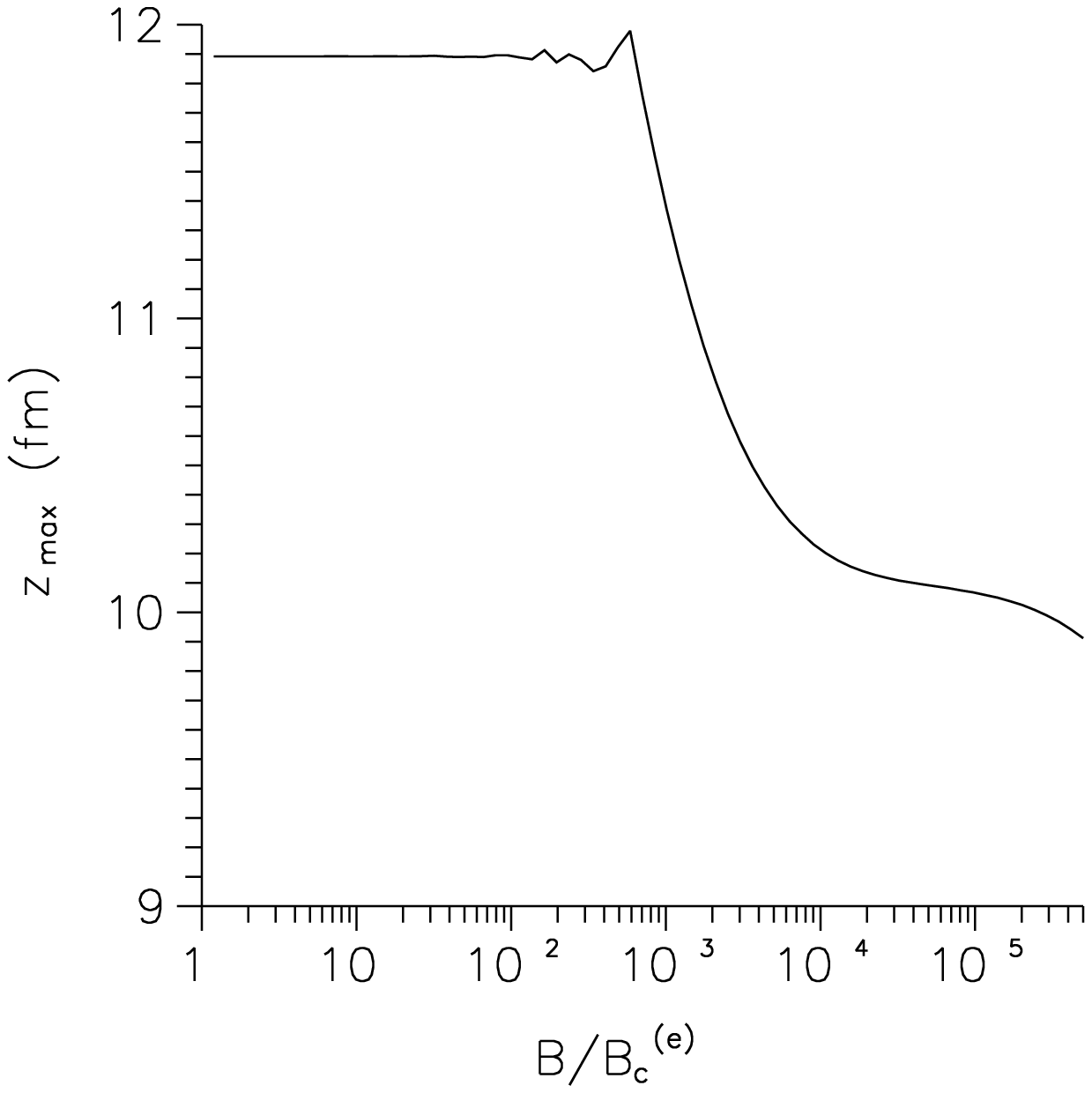}}
\caption{The variation of $z_{max}$ with the magnetic field strength.}
\label{fig:7}
\end{figure*}
\begin{figure*}
\resizebox{0.75\textwidth}{!}{%
\includegraphics{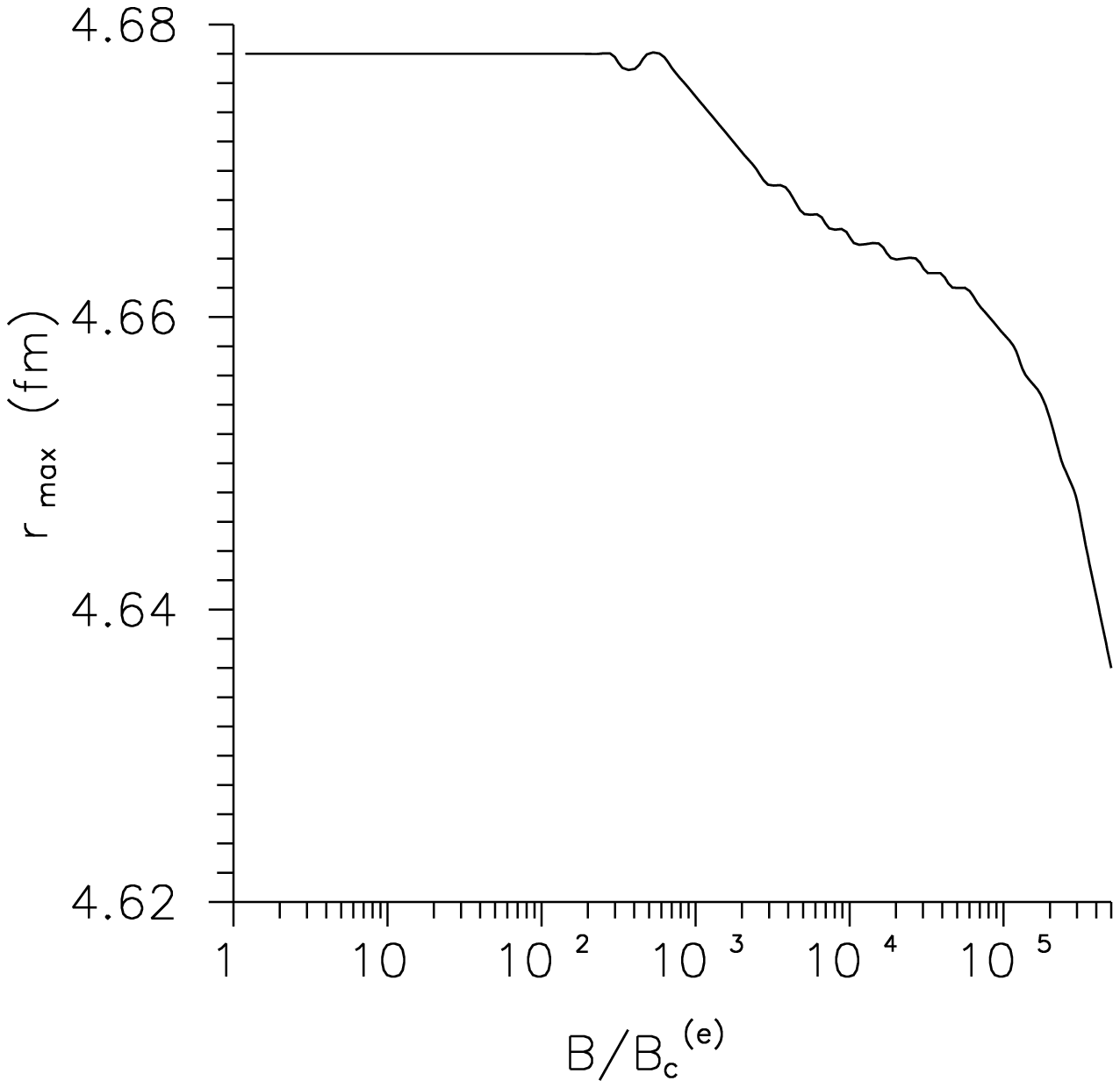}}
\caption{The variation of $r_{max}$ with the magnetic field strength.}
\label{fig:8}
\end{figure*}
\begin{figure*}
\resizebox{0.75\textwidth}{!}{%
\includegraphics{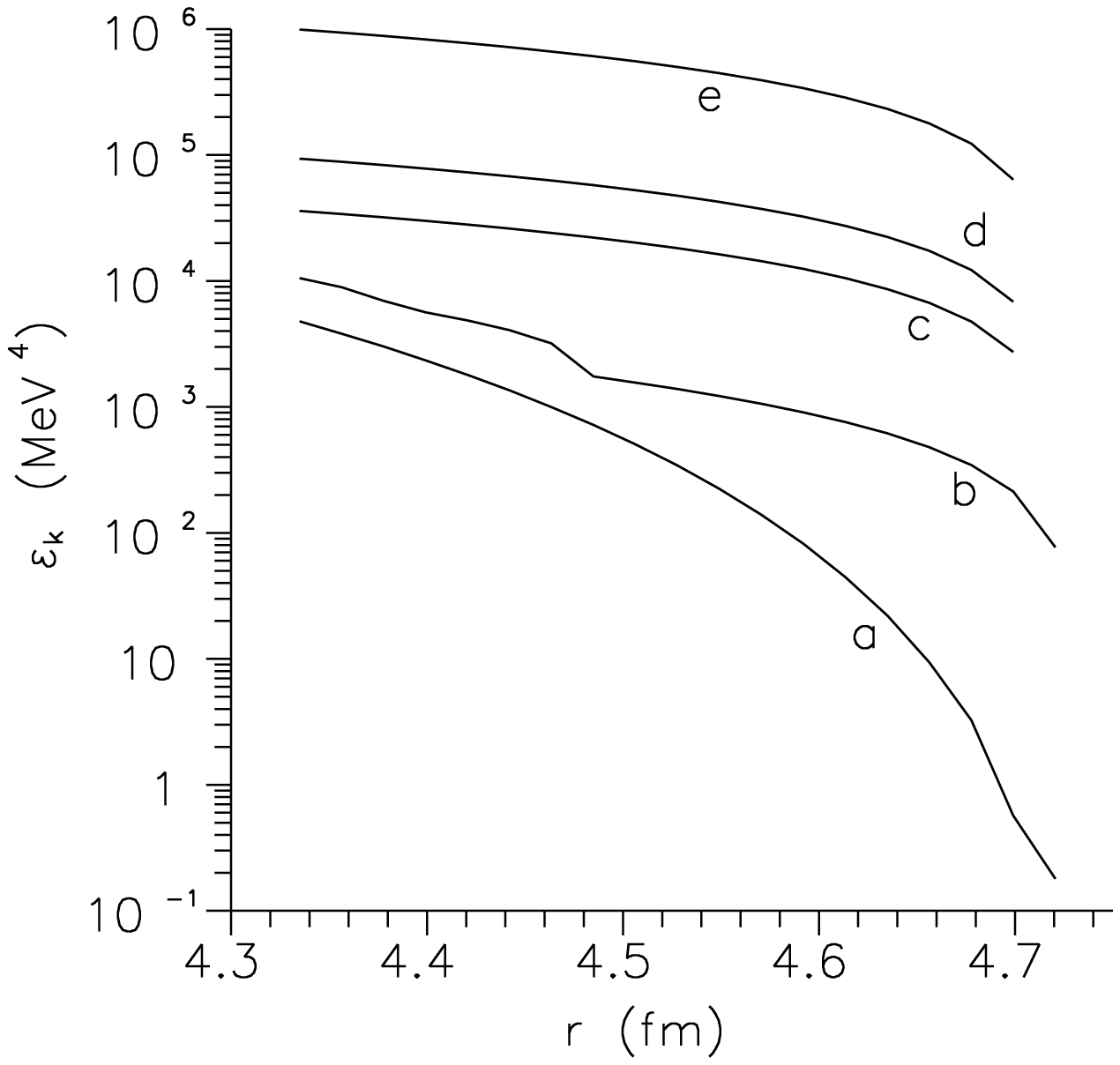}}
\caption{The variation of kinetic energy density with the radial distance in Fermi. The curves
indicated by $a$, $b$, $c$, $d$ and $e$ are for $B=10,
5\times 10^2, 5\times 10^3, 1\times 10^4$ and $5\times 10^4$ times $B_c^{(e)}$ respectively.}
\label{fig:9}
\end{figure*}
\begin{figure*}
\resizebox{0.75\textwidth}{!}{%
\includegraphics{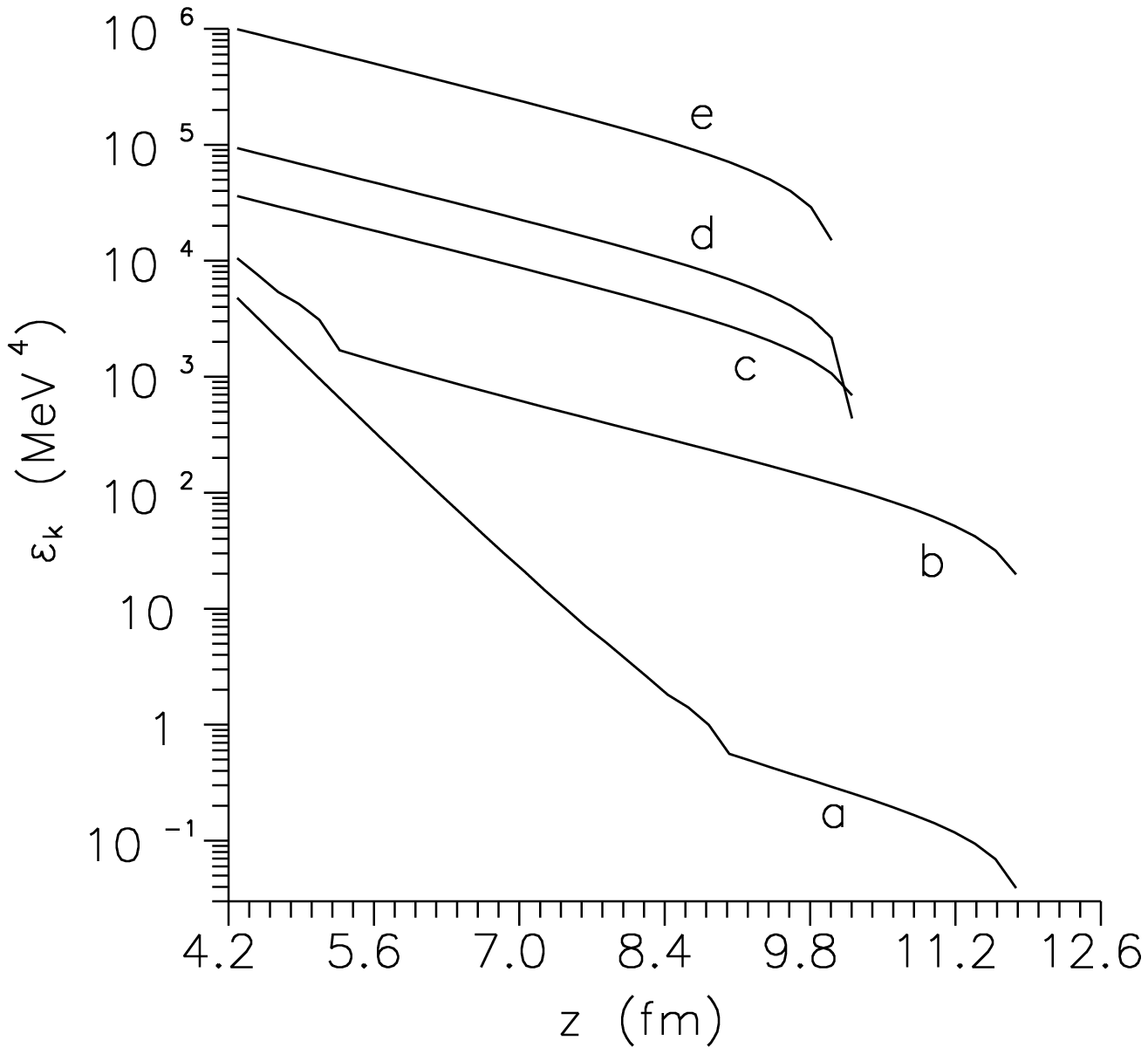}}
\caption{The variation of kinetic energy density with the axial coordinate in Fermi. The curves
indicated by $a$, $b$, $c$, $d$ and $e$ are for $B=10,
5\times 10^2, 5\times 10^3, 1\times 10^4$ and $5\times 10^4$ times $B_c^{(e)}$ respectively.}
\label{fig:10}
\end{figure*}
\begin{figure*}
\resizebox{0.75\textwidth}{!}{%
\includegraphics{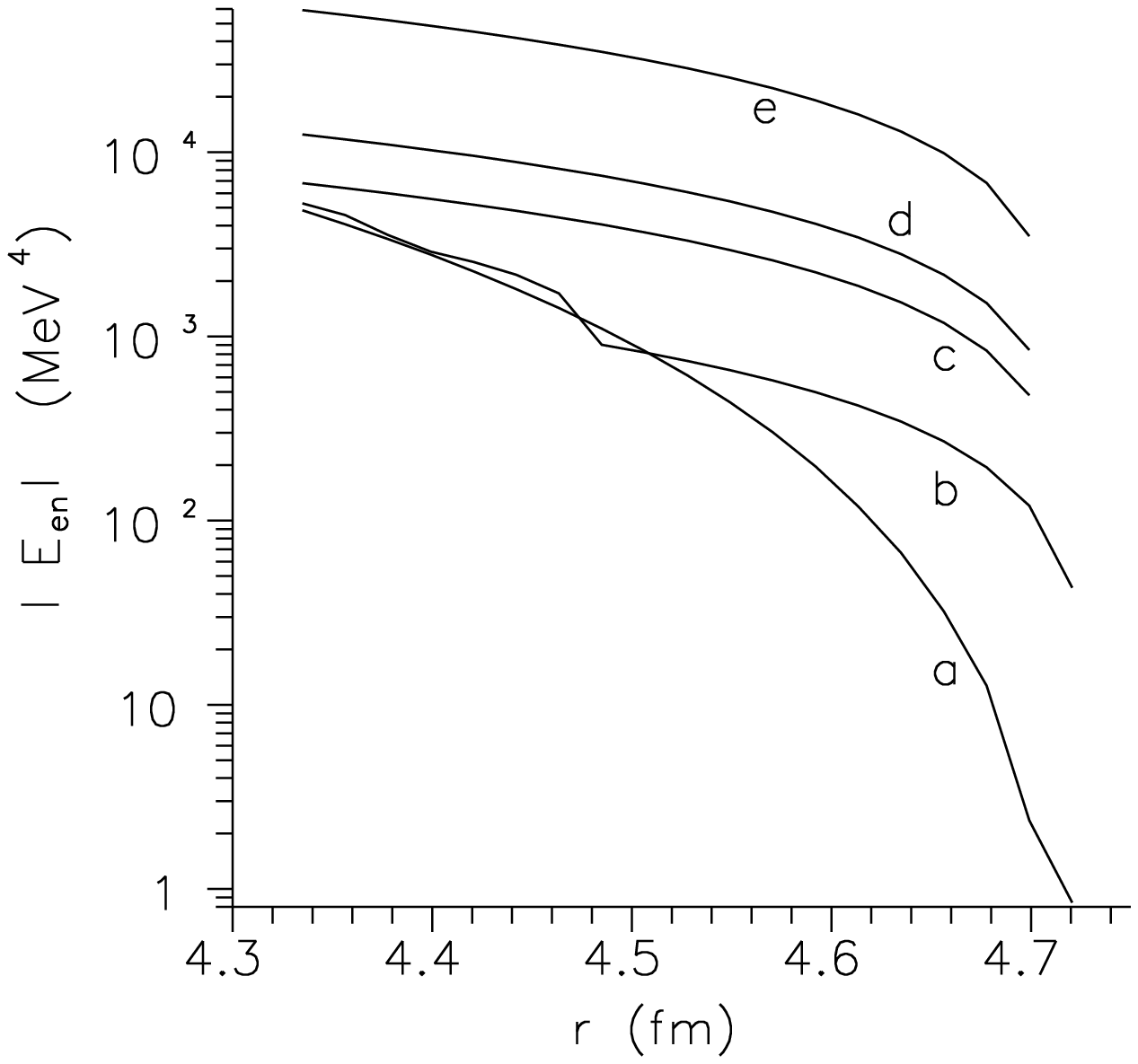}}
\caption{The variation of the magnitude of electron nucleus interaction energy density with the radial 
distance in Fermi. The curves indicated by $a$, $b$, $c$, $d$ and $e$ are for $B=10,
5\times 10^2, 5\times 10^3, 1\times 10^4$ and $5\times 10^4$ times $B_c^{(e)}$ respectively.}
\label{fig:11}
\end{figure*}
\begin{figure*}
\resizebox{0.75\textwidth}{!}{%
\includegraphics{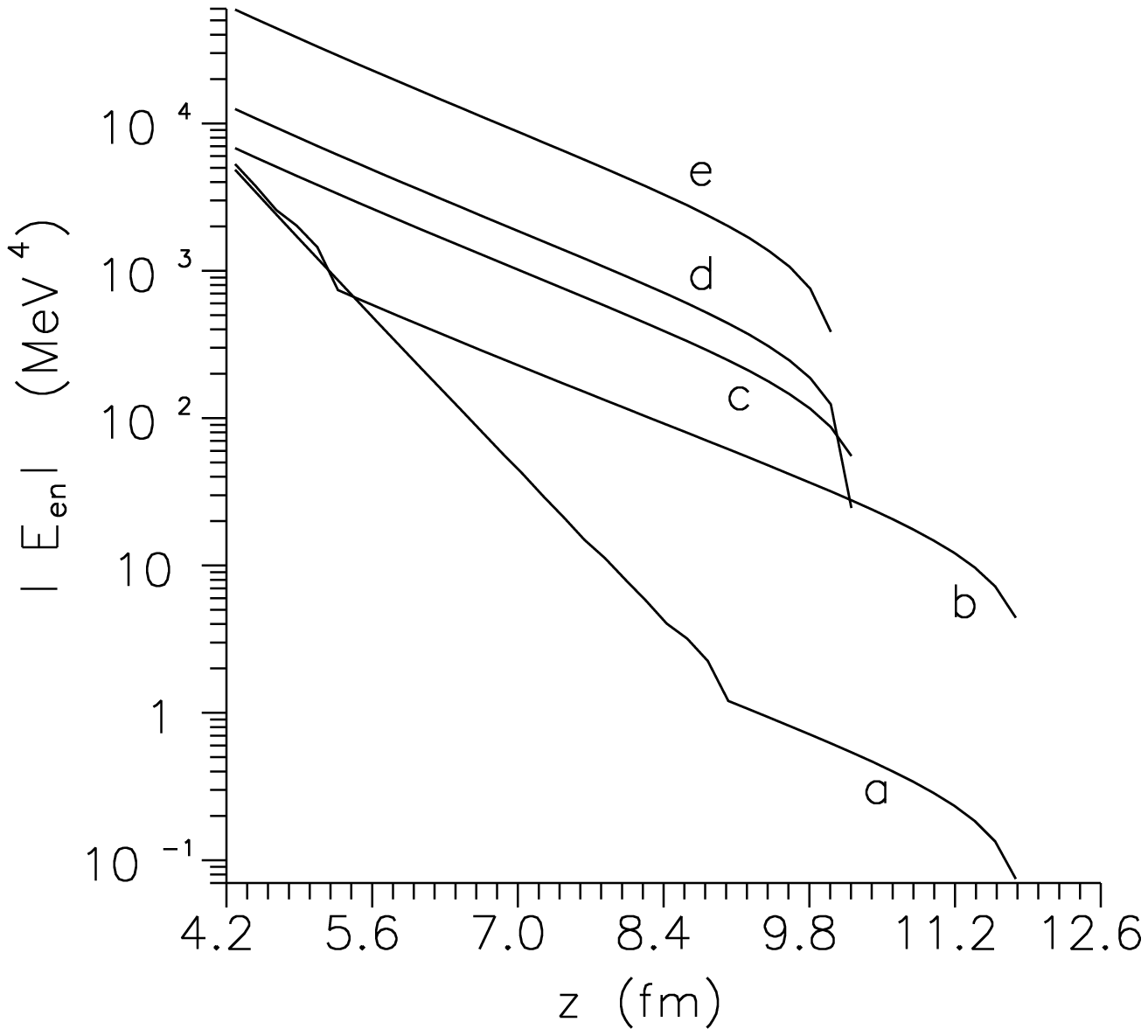}}
\caption{The variation of the magnitude of  electron nucleus interaction energy density with the 
axial coordinate in
Fermi. The curves indicated by $a$, $b$, $c$, $d$ and $e$ are for $B=10,
5\times 10^2, 5\times 10^3, 1\times 10^4$ and $5\times 10^4$ times $B_c^{(e)}$ respectively.}
\label{fig:12}
\end{figure*}
\begin{figure*}
\resizebox{0.75\textwidth}{!}{%
\includegraphics{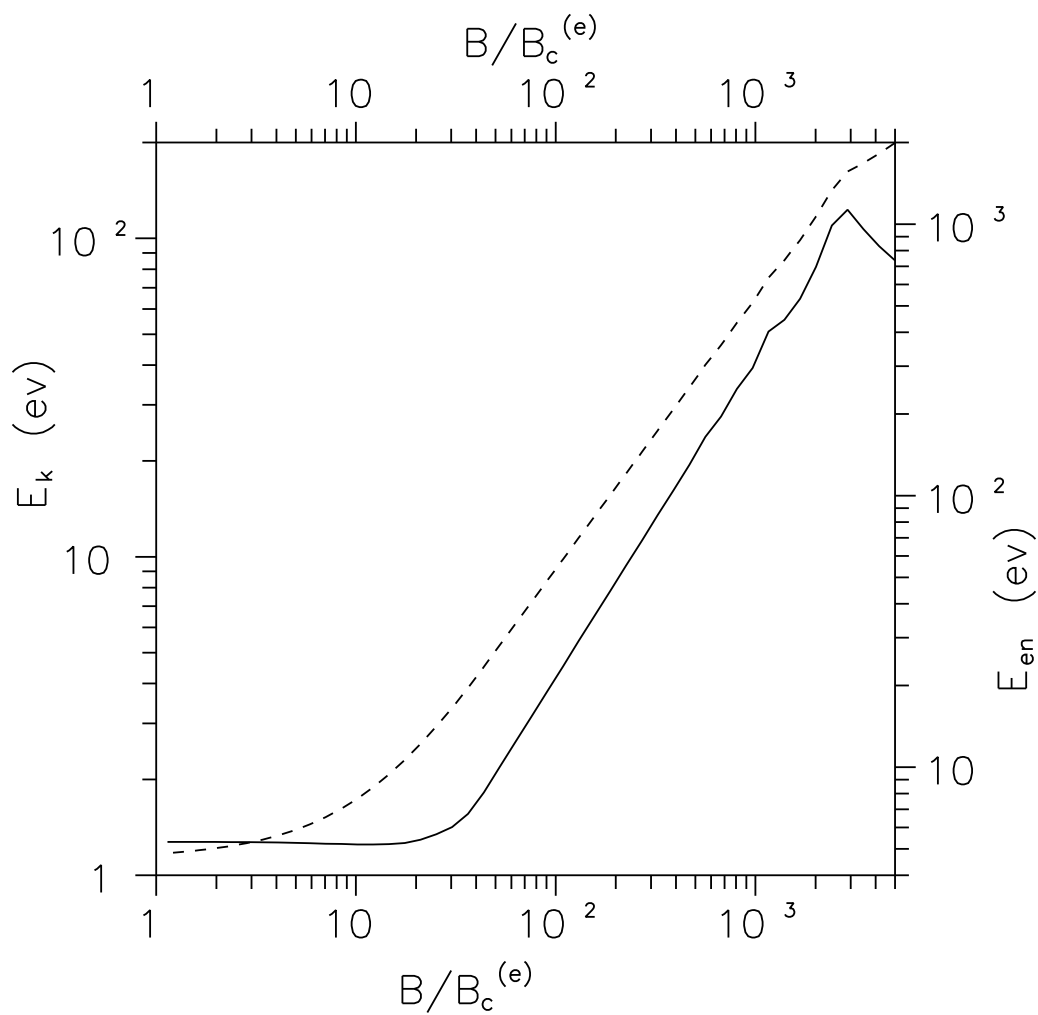}}
\caption{The variation of cell averaged kinetic energy (solid curve) and cell averaged
magnitude of the electron-nucleus interaction energy (dashed curve) with the strength 
of magnetic field.} 
\label{fig:13}
\end{figure*}
\begin{figure*}
\resizebox{0.75\textwidth}{!}{%
\includegraphics{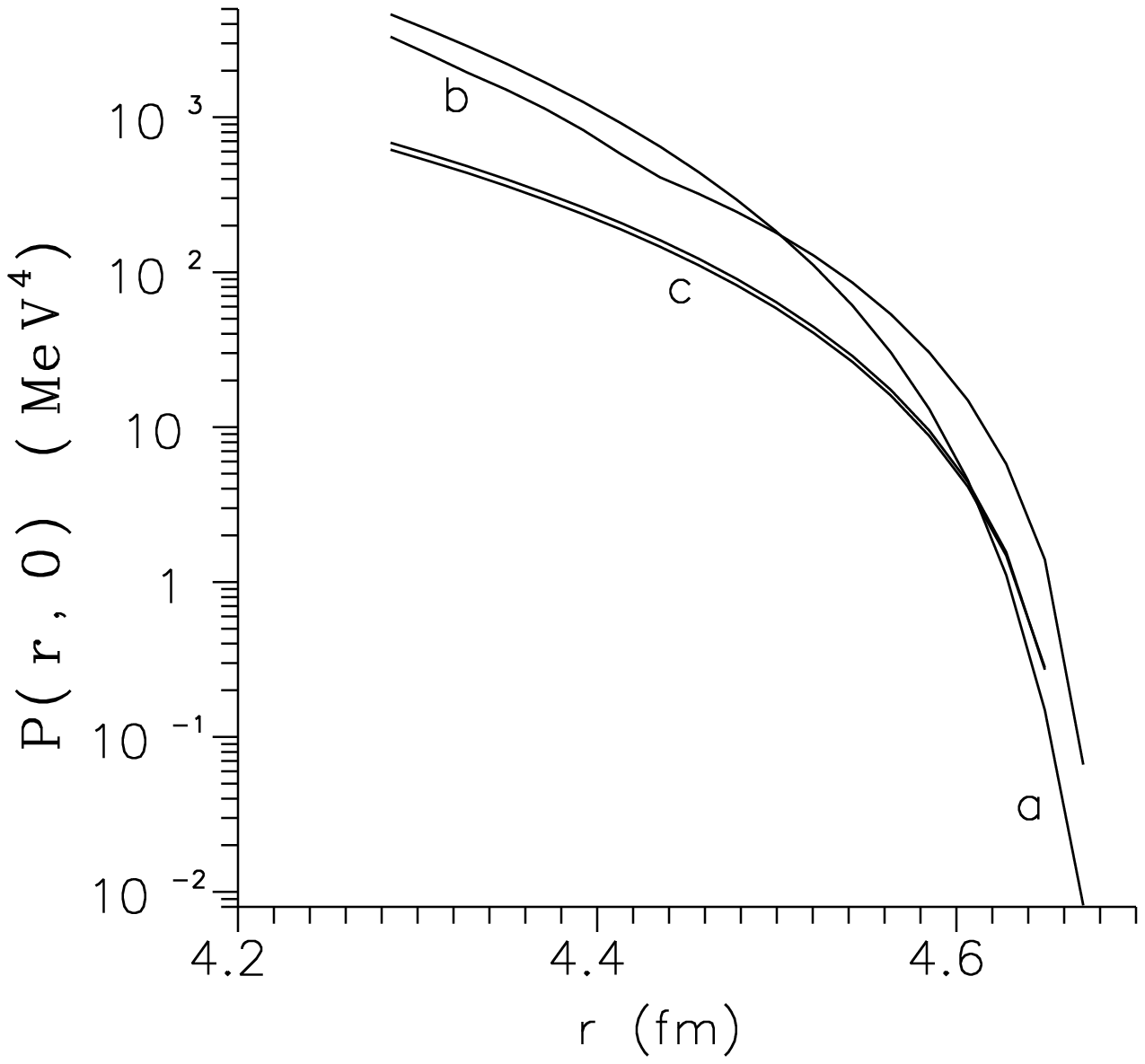}}
\caption{The variation of kinetic pressure of non-uniform electron gas within WS cell
with the radial distance in Fermi. Curves $a$ and $b$ are for $B=10\times B_c^{(e)}$ and $5\times
10^2B_c^{(e)}$ respectively, whereas upper and the lower curves indicated by $c$ are for $5\times
10^3B_c^{(e)}$ and $10^4\times B_c^{(e)}$ respectively.}
\label{fig:14}
\end{figure*}
\begin{figure*}
\resizebox{0.75\textwidth}{!}{%
\includegraphics{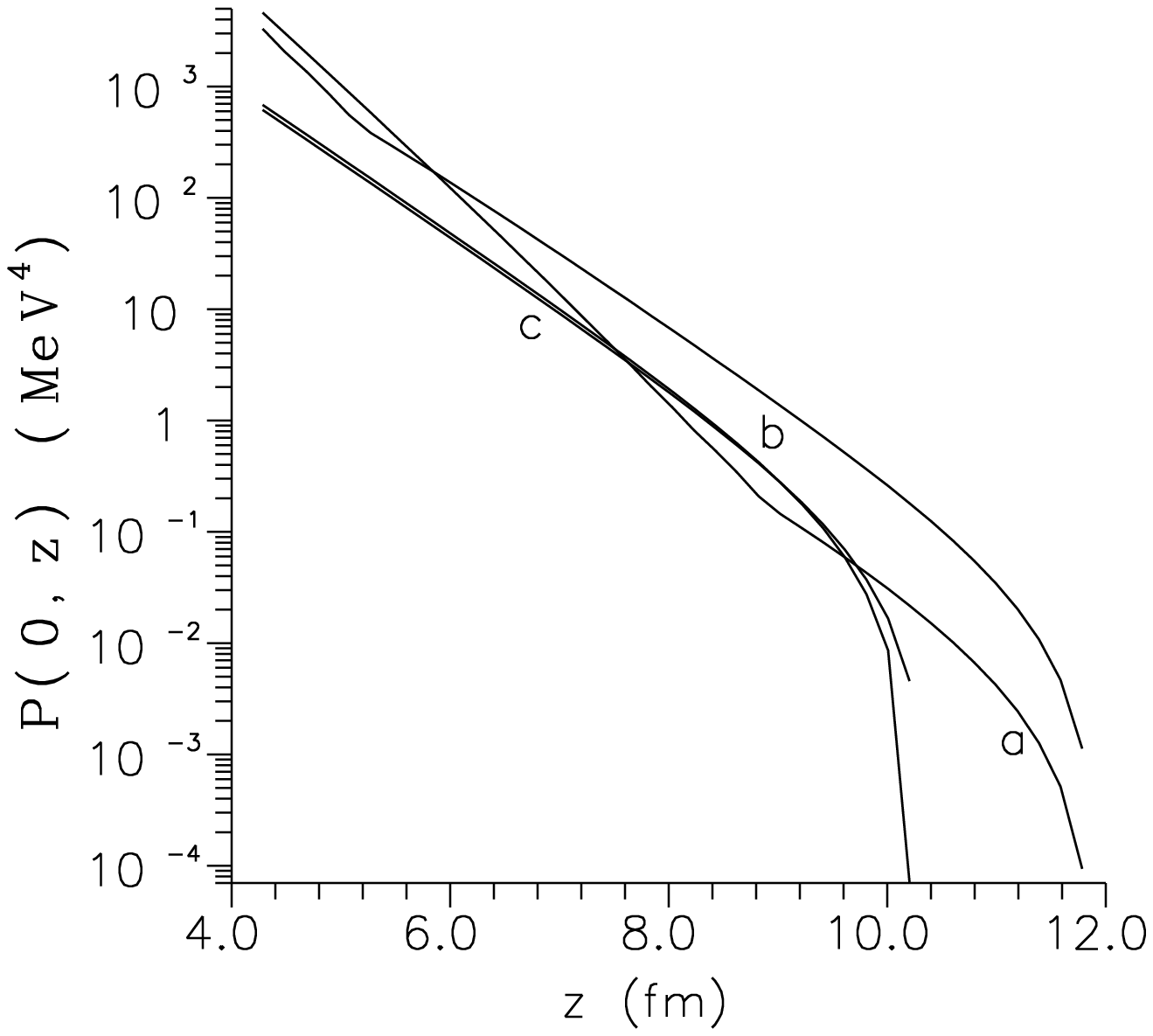}}
\caption{The variation of kinetic pressure of non-uniform electron gas within WS cell with the axial
coordinate in Fermi. Curves $a$ and $b$ are for $B=10\times B_c^{(e)}$ and $5\times
10^2B_c^{(e)}$ respectively, whereas upper and the lower curves indicated by $c$ are for $5\times
10^3B_c^{(e)}$ and $10^4\times B_c^{(e)}$ respectively.}
\label{fig:15}
\end{figure*}
\begin{figure*}
\resizebox{0.75\textwidth}{!}{%
\includegraphics{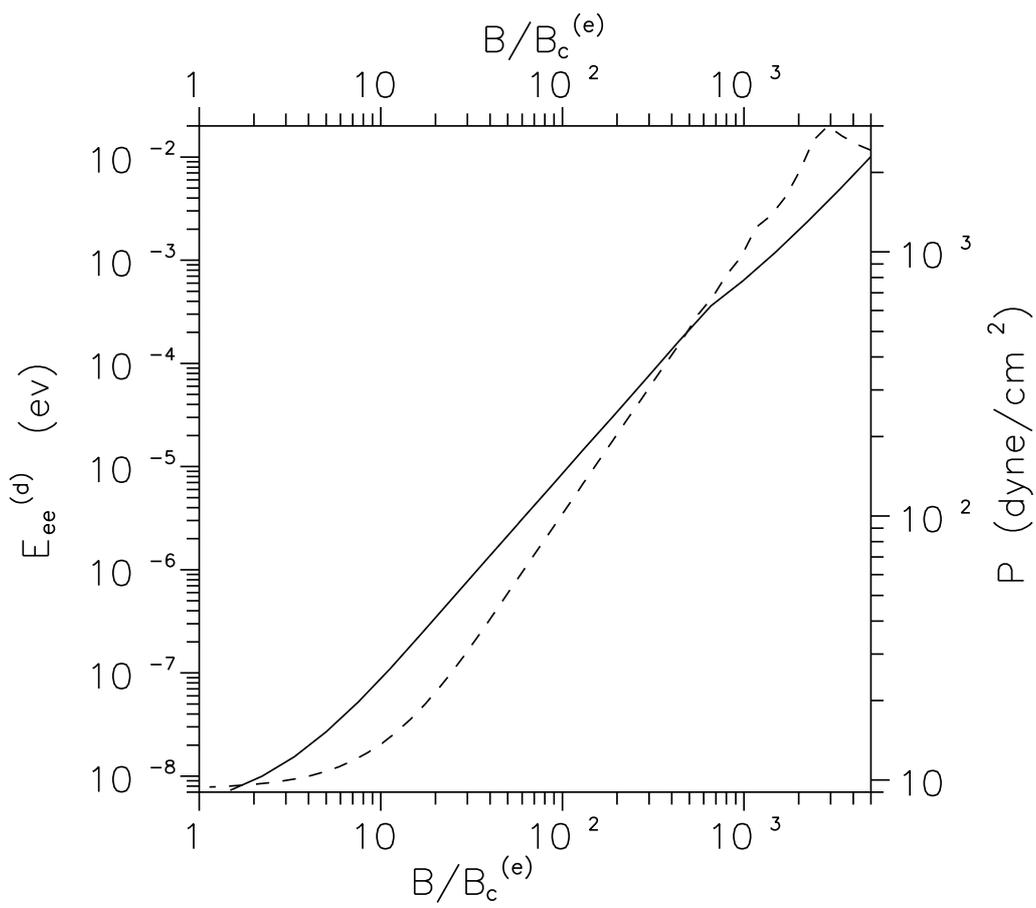}}
\caption{The variation of cell averaged electron-electron direct interaction energy in ev (solid
curve) and the cell averaged electronn kinetic pressure (dashed curve) with the strength of magnetic
field.}
\label{fig:16}
\end{figure*}
\clearpage
\begin{figure*}
\resizebox{0.75\textwidth}{!}{%
\includegraphics{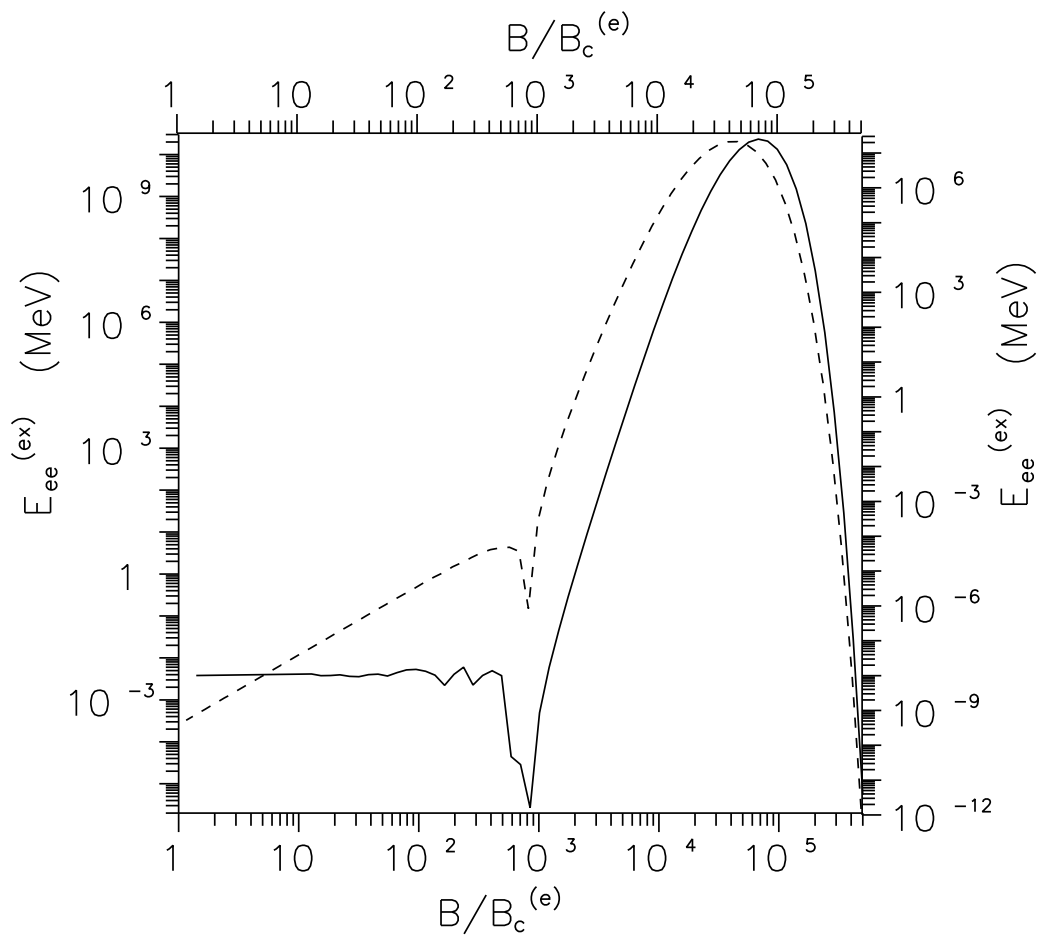}}
\caption{The variation of cell averaged magnitude of the electron-electron exchange energy for 
$\nu \neq 0$ (solid curve) and the same quantity with $\nu=0$ (dashed curve) with the strength of magnetic 
field.} 
\label{fig:17}
\end{figure*}
\begin{figure*}
\resizebox{0.75\textwidth}{!}{%
\includegraphics{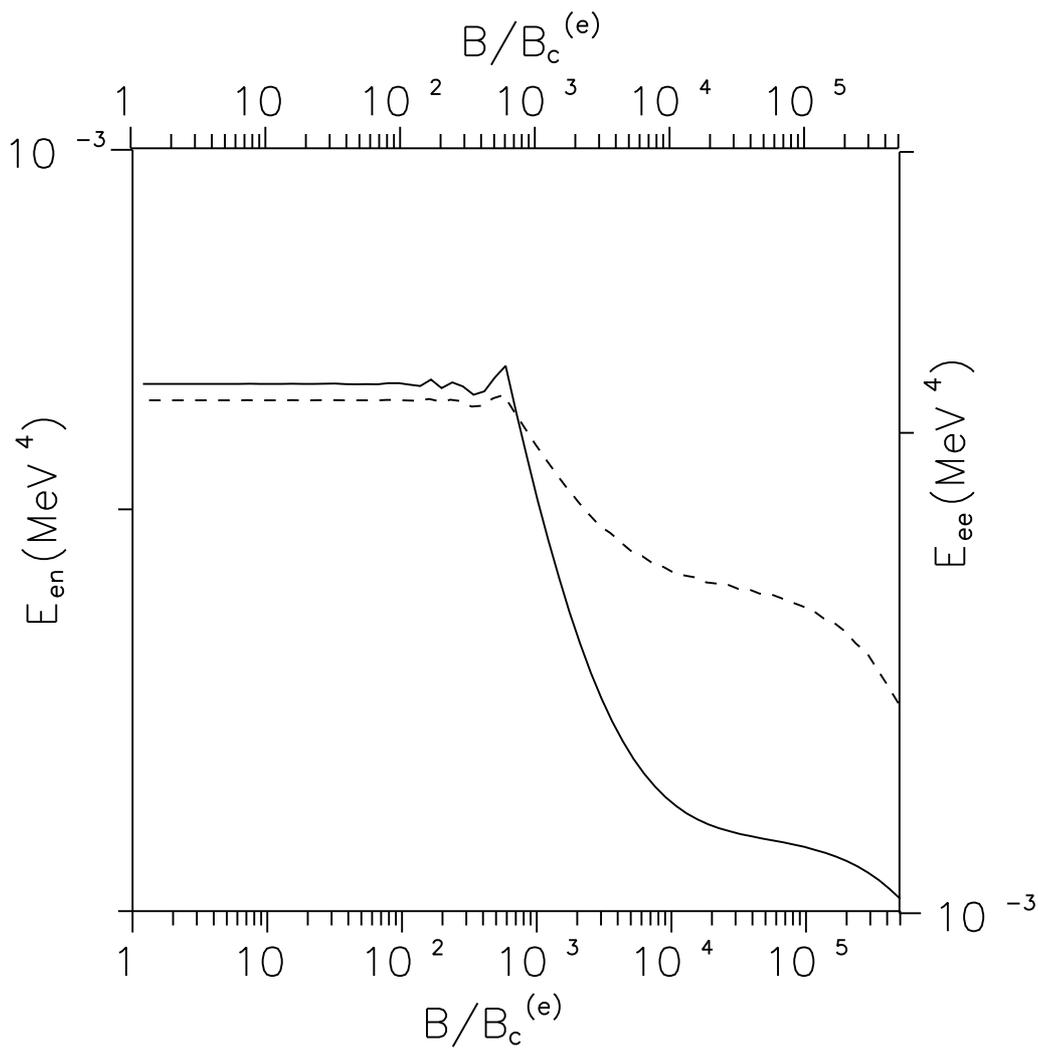}}
\caption{The variation of cell averaged electron-electron Coulomb energy density (solid
curve) and electron-nucleus Coulomb energy density (dashed curve) with the strength of magnetic field.} 
\label{fig:18}
\end{figure*}
\begin{figure*}[ht]
\resizebox{0.75\textwidth}{!}{%
\includegraphics{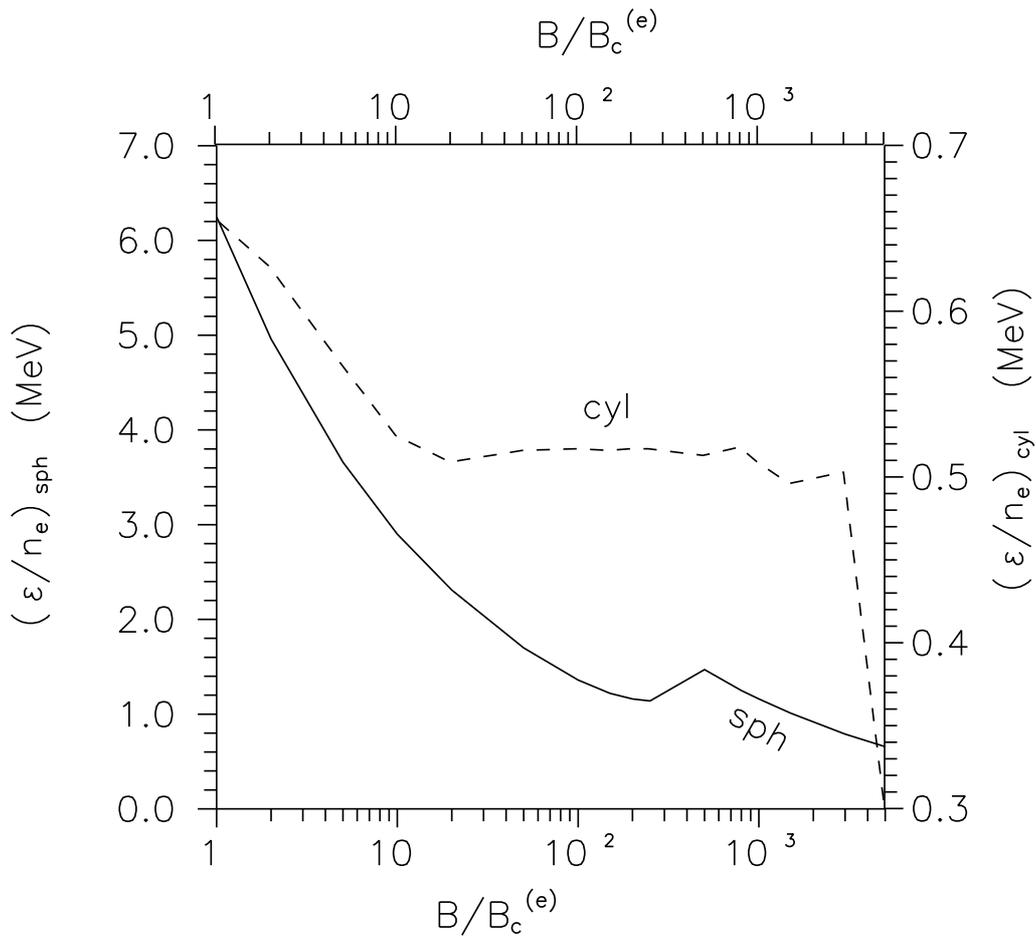}}
\caption{Variation of total energy per electron in MeV with the strength of
magnetic field. The solid curve is for the spherical case with
energy per electron plotted along left y-axis and the dashed one is for 
the cylindrically deformed case with energy per electron plotted along
right y-axis.} 
\label{fig:19}
\end{figure*}
\end{document}